\documentclass{ws-procs9x6}     
\usepackage{wrapfig}

\begin{document}

%\title{For proceedings contributors:\\Using World Scientific's WS-procs9X6\\document class in \LaTeX2e}
\title{Lectures on 
\\
Higher-Gauge Symmetries from Nambu Brackets \\
and \\Covariantized M(atrix) Theory
\footnote{Lectures delivered in the workshop ``Higher Structures in String Theory and M-theory", TFC Thematic Program, {\it Fundamental Problems on Quantum Physics},  Tohoku University (March 7-11, 2016), to be published in the proceedings.}
} 

%\author{A. B. Author$^*$ and C. D. Author}

%\address{University Department, University Name,\\
%City, State ZIP/Zone, Country\\
%$^*$E-mail: ab\_author@university.com\\
%www.university\_name.edu}

\author{T. Yoneya}

\address{Institute of Physics,
Komaba, University of Tokyo,\footnote{
Emeritus Professor}
 Japan\\
%E-mail: tam@hep1.c.u-tokyo.ac.jp
}
\begin{abstract}
This lecture consists of three parts. In part I, an overview is given on 
the so-called 
 Matrix theory in the light-front gauge as a 
proposal for a concrete and non-perturbative formulation of M-theory. I emphasize motivations towards 
its covariant formulation. Then, in part II, I turn the 
subject to the so-called Nambu bracket and Nambu mechanics,  
which were proposed by Nambu in 1973 as a possible extension 
of the ordinary Hamiltonian mechanics. After reviewing 
briefly Nambu's original work, it will be explained why his idea 
may be useful in exploring higher symmetries which would 
be required for covariant formulations of Matrix theory. Then, 
using this opportunity, 
some comments on the nature of Nambu mechanics and 
its quantization are given incidentally: though they are 
not particularly relevant for our specialized purpose of 
constructing covariant Matrix theory, they may be of 
some interests for further developments in view of possible 
other applications of Nambu 
mechanics. The details will be relegated to forthcoming publications. 
In part III, I give an expository account of the basic ideas and main results from my
 recent attempt to 
construct a covariantized Matrix theory on the basis of 
a simple matrix version of Nambu bracket equipped with 
some auxiliary variables, which characterize the scale of 
M-theory and simultaneously play a crucial role in 
realizing (dynamical) supersymmetry in a covariant fashion.

\end{abstract}

\keywords{M-theory, M(atrix) theory, Nambu mechanics, 
Nambu bracket.}

\bodymatter

\newpage
\begin{center}
\vspace{0.3cm}
Part I : An Overview on Matrix Theory
\vspace{-0.5cm}
\end{center}

\section{The M-theory conjecture}\label{aba:sec1}
M-theory was conjectured in the mid 90s as a 
hidden theory: it would play a crucial pivotal 
role in a possible non-perturbative formulation 
unifying five perturbative string theories which 
had been established in the mid 80s. 
The basic tenets of M-theory are as follows:
\begin{enumerate}
\item It achieaves a complete unification of strings 
and D-branes in a compactified $(10,1)$ dimensional spacetimes. 
\item There is a unique fundamental length scale $\ell_{11}$ corresponding 
to the 11 dimensional Planck scale. Together with the radius $R_{11}$ of 
compactification of (10,1)-dimensional 
spacetime to $(9,1)$ dimensional spacetimes 
corresponding to type IIA string theory (and 
$E_8\times E_8$ heterotic string theory), it sets the string scale $\ell_s$ and string coupling constant 
$g_s^{(A)}$ as
\begin{align}
\ell_s^2=\ell_{11}^3R_{11}^{-1}, \qquad 
g_s^{(A)}=(R_{11}\ell_{11}^{-1})^{3/2}=R_{11}/\ell_s.
\nonumber 
\end{align}
The scales and couplings of the other perturbative string theories are 
related by duality relations. For instance, the so-called 
S-duality of type IIB theory is 
explained by introducing additional compactification along 
one of remaining spatial directions with radius $R_{10}$: 
The type IIA and IIB theories are then related by a 
 T-duality transformation, 
\begin{align}
R_{10}\rightarrow \ell_s^2/R_{10}, \quad 
g_s^{(A)}\rightarrow g_s^{(B)}=g_s^{(A)}\ell_s/R_{10}=R_{11}/R_{10}. 
\nonumber
\end{align}
Thus the S-duality transformation $g_s^{(B)}\rightarrow 1/g_s^{(B)}$ of type IIB theory corresponds simply to the interchange of 10th and 11th directions, 
$(R_{10},R_{11})\leftrightarrow (R_{11},R_{10})$. 

\item The underlying dynamical degrees of freedom are 
super-membranes (or M2-banes) which have an ``electrical charge" 
coupled to a 3-form gauge field as particular components of 
physical degrees of freedom of super-membranes. 
There are also excitations, called M5-branes, which correspond to 
 excitations which are electro-magnetically dual to super-membranes. 
After compactification to $(9,1)$ dimensions, the super-membranes 
behave either as fundamental strings when one of their two 
spatial directions is wrapped 
along the compactified direction, or as D2-branes when none of the spatial directions of supermembranes are wrapped along the compactified direction. 
\end{enumerate}
In particular, as a consequence of these assumptions, 
gravitons in compactified 11 dimensions, if momentum 
along the compactified direction is zero, are the ground-state modes 
of strings at least in the limit of small 
compactifcation radius, and if they have non-zero momentum along the 
compactified direction, are the Kaluza-Klein modes 
which should coincide with the ground-state of 
wrapped super-membrane with non-zero momentum 
in the same direction and are 
identified with D0-branes (or D-particles) of 
type IIA string theory. 
This picture is valid for small compactification radius 
$R_{11}\ll \ell_{11}$ with fixed $\ell_s=\ell_{11}/(R_{11}/\ell_{11})^{1/2}
=\ell_{11}g_s^{-1/3}$. The latter relation shows that, 
in the opposite limit of de-compactification $R_{11}\gg \ell_{11}$ 
with fixed $\ell_{11}$, we have $\ell_s\rightarrow 0$ and 
$g_s^{(A)}\rightarrow \infty$, namely, 
a very singular limit of type IIA string theory corresponding to 
{\it infinite} string tension and {\it infinite} string coupling. 
Since string theory has been known only perturbatively 
in the limit of infinitely small $g_s$, it is very difficult to 
imagine how such a peculiar limit should be formulated. 
One suggestive expectation is that M-theory might be described 
by some degrees of freedom corresponding to 
short strings and its KK modes or D0-branes but with some intrinsic non-perturbative 
interactions among them, which would make possible some 
mechanisms generating not only supermembranes, but also other physical degrees of 
freedom as some sort of bound states of D0-branes (and short strings). 

\section{The dynamics of (Super)membranes}
A similar picture which seems to be 
compatible with the foregoing viewpoint naturally emerges itself
 if we envisage 
the dynamics of super-membranes. 
To study the relativistic dynamics of membranes 
assuming flat 11-dimensional spacetimes, 
we can start from a typical action 
\begin{align}
A=-\frac{1}{\ell_{11}^3}
\int d^3 \xi \Bigl(\frac{1}{e}\{X^{\mu},X^{\nu},X^{\sigma}\}
\{X_{\mu},X_{\nu},X_{\sigma}\}-e\Bigr)+\ldots
\label{membaction}
\end{align}
where $X^{\mu}(\xi)$ ($\mu, \nu, \dots, \in (1,2,\ldots, 10, 0)$) are 
the target space coordinates of the membrane and the ellipsis ($\ldots$) means other contributions involving 
in particular the fermionic degrees of freedom.  
Throughout this lecture, we always 
use Einstein's summation convention 
for spacetime (and/or space) indices in target 
space.  The variable $e=e(\xi)$ is 
an auxiliary field, transforming as a world-volume density 
under 3-dimensional diffeomorphism 
with respect to the parameterization $(\xi^1,\xi^2, \xi^0)
=(\sigma_1,\sigma_2, \tau)$ of the world-volume of a membrane. 
We also used the following notation,  
\begin{align}
\{X^\mu,X^\nu, X^\sigma\}\equiv \sum_{a,b,c\in (1,2,0)}\epsilon^{abc}
\partial_{a}X^\mu\partial_{b}X^{\nu}\partial_{c}X^{\sigma}, 
\label{NPbracket}
\end{align}
which 
will be called ``Nambu-bracket" (or Nambu-Poisson bracket). 
The standard form (Dirac-Nambu-Goto type) of the world-volume action 
is obtained by eliminating the auxiliary field $e$. 

Unfortunately, this is a notoriously difficult system to deal with, 
especially with respect to quantization. 
Only tractable way which allow us a reasonably concrete treatment so far is to 
adopt the light-front gauge $X^+\equiv X^{10}+X^0=\tau$, breaking 
11-dimensional Lorentz covariance.\cite{berg} After a further (still partial) gauge-fixing of the residual (time-dependent) 
re-parametrizations of spatial coordinates $(\sigma_1, \sigma_2)$ 
by demanding that the induced metric of the world-volume takes an 
orthogonal form $ds^2=g_{00}d\tau^2+g_{rs}d\sigma^rd\sigma^s$ 
($r,s\in (1,2)$) with $g_{rs}=\partial_rX^i\partial_sX^i$ 
and also that light-like momentum density is a constant $P^+
=P^{10}+P^0$ 
with the normalization $\int d^2\sigma=1$, we 
are left with a constraint 
\begin{align}
\{P_i, X^i\}+\cdots \approx 0, 
\nonumber 
\end{align}
where \begin{align}
\{A, B\}
\equiv \partial_1A\partial_2B-\partial_2A\partial_1B
\nonumber 
\end{align}
for arbitrary pair of functions $A, B$, 
and the effective Hamiltonian, 
in the unit $\ell_{11}=1$ for notational brevity :
\begin{align}
H=\int d^2\sigma\frac{1}{P^+}(P_i^2+\frac{1}{2}
\{X^i, X^j\}^2)+\cdots.
\nonumber 
\end{align}
where the indices $i, j, \ldots, $ of 
the target-space coordinates $X^i$ run over only 
SO(9) transverse directions $(1,2,\ldots, 9)$. 
%The ellipses $(\cdots )$denotes other contributions including fermionic variables. 
The above 
constraint demands that the system is invariant under
infinitesimal (time-independent) area-preserving 
diffeomorphism of spatial coordinates which still remains as 
residual gauge symmetry after all of the above 
gauge-fixing conditions:
\begin{align}
\delta_F X^i=(\partial_1F\partial_2-\partial_2F\partial _1)X^i
=\{F, X^i\}
\label{areapresdiff}
\end{align}
with $F=F(\sigma)$ is an arbitrary function of the spatial 
world-volume coordinates.

As a 3-dimensional field theory, this is still a very nontrivial 
system without standard kinetic-potential terms, 
such as $(\partial_1X^i)^2+(\partial_2X^i)^2$, of 
second order, but being instead equipped with (non-renormalizable) quartic interaction terms with 
four derivatives. A useful suggestion for controlling this system 
was made by Goldstone and Hoppe \cite{goldhoppe} in the early 80s (and developed further in ref. \cite{dhn} later). Namely, 
we can regularize this system by replacing the fields $X^i(\xi)$ 
by finite $N\times N$ hermitian matrices $X^i_{ab}$ 
where the matrix indices $a,b, \ldots$ now run 
from 1 to $N$. Then, the above Hamiltonian is replaced by
\begin{align}
H=\frac{1}{P^+}{\rm Tr}\Bigl(\boldsymbol{P}_i^2-\frac{1}{2}
[\boldsymbol{X}^i, \boldsymbol{X}^j]^2\Bigr)+\cdots ,
\label{memham}
\end{align}
where and in what follows we use slanted boldface symbols $\boldsymbol{X}^i, 
\boldsymbol{P}^i$ (hence, $(\boldsymbol{X}^i)_{ab}\equiv X^i_{ab}$) for  matrices when the matrix indices $(a, b, \ldots)$ are 
supressed. $\boldsymbol{P}_i$'s are of course 
canonical-momentum matrices corresponding to the 
canonical-coordinate matrices $\boldsymbol{X}^i$'s. 
The constraint corresponding to area-preserving 
diffeomorphism 
is now replaced by 
\begin{align}
[\boldsymbol{P}_i,\boldsymbol{X}^i]
+\cdots \approx 0, 
\label{matrixconstaint}
\end{align}
which generates infinitesimal 
unitary (SU($N$)) transformations of matrix variables:
\begin{align}
\delta_F\boldsymbol{X}^i=i[\boldsymbol{F}, \boldsymbol{X}^i], 
\quad 
\delta_F\boldsymbol{P}_i=i[\boldsymbol{F}, \boldsymbol{P}_i]
\nonumber 
\end{align}
where $\boldsymbol{F}$ is an arbitrary (time-independent) 
hermitian matrix. 

It should be clear that the matrix 
regularization is based on a formal but natural 
analogy between classical brackets $\{\,\, , \,\, \}$ and 
commutator $i[\,\, , \,\, ]$.\footnote{Such an analogy had previously been 
suggested by Nambu\cite{nambu1} 
in string theory, 
in connection with the so-called Schild action which can actually 
be regarded as the string version of the action \eqref{membaction} 
in the gauge $e=1$.}
The basic idea here is that given a finite 
world-volume with fixed topology we can 
alway use some appropriate Fourier-like representation 
for the fields $X^i(\sigma), P_i(\sigma)$ and take the resulting 
discretized Fourier components of them as dynamical variables. 
If we have an appropriate way of truncating the 
infinite number of such Fourier components into 
a finite set of components by keeping  
the remnant of the area-preserving 
diffeormorphism group as a symmetry group of this finite set, 
it would provide a desirble regularized version of 
the original system.  It is not unreasonable to
 expect that, for sufficiently large $N$, 
the above system would be capable of approximating 
arbitrary kinds of fixed topology of supermembrane in some 
classical limit and, in quantum theory, of describing 
the dynamics of supermembranes and other 
physical objects. Now with a finite number of degrees of 
freedom, the system is completely well defined and therefore 
amenable to any non-perturbative studies including 
computer simulations. Matrix models of this kind would play, at the very least, the role of the same sort 
that lattice gauge theories are playing in non-perturbative studies of gauge field theories. The importance of such tractable system in this sense should not be 
underestimated in view of the genuine dynamical 
nature of the M-theory conjecture.

\section{M(atrix) theory 
proposal in the DLCQ scheme}
One of various remarkable facts concerning the matrix regularization of supermembrane is 
that the same system appears as the low-energy effective theory\cite{witten} 
of $N$ D0-branes. In the same unit $\ell_{11}=1$ ($\ell_s=g_s^{-1/3}, 
R_{11}=\ell_s^{-2}=g_s^{2/3}$) as above, the 
Hamiltonian is 
\begin{align}
H_{{\rm D0}}=\frac{1}{2}g_s^{2/3}{\rm Tr}\Bigl(\boldsymbol{P}_i^2
-\frac{1}{2}[\boldsymbol{X}_i, \boldsymbol{X}_j]^2\Bigr)+\cdots, 
\label{d0hamiltonian}
\end{align} 
where the momentum is given as
\begin{align}
\boldsymbol{P}_i=\partial_0\boldsymbol{X}_i+i[\boldsymbol{A}, \boldsymbol{X}^i]
\nonumber 
\end{align}
with $\boldsymbol{A}$ being an SU($N$) gauge matrix-field 
corresponding to local 
gauge transformation
\begin{align}
&\delta \boldsymbol{X}_i=i[\boldsymbol{F}, \boldsymbol{X}_i], 
\label{gaugex}
\\
&\delta \boldsymbol{A}=i[\boldsymbol{F}, \boldsymbol{A}]-
\frac{d\boldsymbol{F}}{dt}.
\label{gaugea}
\end{align}
In this case, the constraint \eqref{matrixconstaint} appears 
as the Gauss constraint corresponding to this 
local gauge symmetry. Thus the original 
diffeomorphism symmetry is now interpreted as a 
local gauge symmetry. 
It should be noted that the gauge field $\boldsymbol{A}$ does 
not play any dynamical role other than giving 
the Gauss constraint, since 
the present system is only (0,1)-dimensional as a gauge field 
theory.

The diagonal components $(X_{i\, aa}, P_{i\, aa})$ of the matrices are interpreted to represent 
the motion of D0-branes, whereas 
the off-diagonal components are supposed to correspond to 
the lowest dynamical degrees of freedom of open strings 
connecting them. 
Thus the zero-mode kinetic energy of D0-branes is 
$\frac{1}{2}g_s^{2/3}\sum_{a=1}^N(P_{i\, aa})^2$.  
This coincides with \eqref{memham} if we assume $P^+=2g_s^{-2/3}=2R_{11}^{-1}$. The latter identification is consistent with 
the assumption that D0-brane is a Kaluza-Klein mode with a 
unit quantized momentum along the compactified
 circle of radius $R_{11}$: if 
$R_{11}$ is sufficiently small,  then we have
\begin{align}
P_{10}=R_{11}^{-1}, \quad 
P^0=\sqrt{P_i^2+P_{10}^2}\rightarrow R_{11}^{-1},
\nonumber 
\end{align}
and hence $P^+\rightarrow 2R_{11}^{-1}$ 
for each independent D0-brane. This is the 
limit where we can trust the above effective low-energy 
Hamiltonian for D0-branes (of mass $1/R_{11}$) in 
weak-coupling string theory in un-compactified $(9,1)$ dimensions. 
Note that if we separate out the center-of-mass momentum $P_{\circ}^i$ 
and the traceless part of the matrices
\begin{align}
P_{\circ}^i\equiv {\rm Tr}\boldsymbol{P}^i, 
\nonumber 
\end{align}
we can write 
\begin{align}
&H=\frac{1}{2P_{\circ}^{10}}(P_{\circ i}^2+\hat{H}), 
\quad 
P_{\circ}^{10}=NR_{11}^{-1}, 
\nonumber \\
&\hat{H}=N{\rm Tr}\Bigl(\hat{\boldsymbol{P}}_i^2
-\frac{1}{2}[\boldsymbol{X}^i, \boldsymbol{X}^j]\Bigr)+\cdots , 
\label{lfhamiltonian}\end{align}
where and in what follows we denote the traceless 
part of the matrices by putting $\,\hat{}\,$ symbol:
$\hat{\boldsymbol{P}}_i=\boldsymbol{P}_i-\frac{1}{N}P_{\circ i}$. $\hat{H}$ involves only the traceless part of the 
matrix variables.

Now what should be the interpretation of the above coincidence? 
Suppose that we consider an ordinary relativistic 
system of $N$ interacting particles in flat 
spacetime. If we extract the center-of-mass momentum 
$P^{\mu}_{\circ}$, the system would have invariably a mass-shell constraint of the form
\begin{align}
P^{\mu}_{\circ}P_{\circ\mu}+M_{{\rm eff}}^2=0, 
\label{covmassshell}
\end{align}
where $M^2_{{\rm eff}}$ is the effective squared mass 
which describes internal (Lorentz-invariant) dynamics 
of the whole system. Using the light-like components, 
this can be expressed as
\begin{align}
-P_{\circ}^-\equiv P_{\circ}^0-P_{\circ}^{10}=
\sqrt{P_{\circ i}^2+(P_{\circ}^{10})^2 +M_{{\rm eff}}^2}-P_{\circ}^{10}\rightarrow 
\frac{P_{\circ i}^2+M_{{\rm eff}}^2}{2P_{\circ}^{10}}
\label{IMF}
\end{align}
in the limit of large $P_{10}$, which corresponds 
to the so-called infinite momentum frame (IMF). 
Alternatively, we can 
use an exact relation using light-like components, irrespectively of $P_{10}$ being large or small, 
\begin{align}
-P^-_{\circ}=\frac{P_{\circ i}^2 +M_{{\rm eff}}^2}{P^+_{\circ}},
\label{DLCQ}
\end{align}
which of course reduces to \eqref{IMF} in the limit 
$P_{\circ}^{10}\rightarrow \infty$. 
In the case of \eqref{DLCQ}, we can assume further that 
the compactification is made directly along the light-like direction $X^-$ with radius $R$ corresponding to the 
quantization condition
\begin{align}
P_{\circ}^+=2N/R, 
\nonumber 
\end{align}
by which \eqref{DLCQ} coincides with \eqref{lfhamiltonian} 
if we identify $\hat{H}$ with $M^2_{{\rm eff}}$. 
This special compactification scheme along $X^-$ is known as the 
discrete light-cone quantization (DLCQ) in field theories. 
But we are now adopting this scheme 
to relativistic system of $N$ particles in 
configuration-space formulation, instead of relativistic 
local field theory where some subtleties are known with 
respect to its significance in non-perturbative properties of 
field theories. 

A crucial difference of this interpretation from 
the IMF is that we can freely change the value of $R$ from infinitely  small to infinitely large, by merely changing the Lorentz frame 
with any {\it fixed} $N$.  In the IMF interpretation, on the contrary,
the radius $R_{11}$ of compactification is fixed irrespectively 
of which Lorentz frame we are studying the system: 
thus for large $P^{10}_{\circ}=N/R_{11}$ we have to take 
large $N$ by assuming that each D-particle has 
fixed eleven-th momentum $1/R_{11}$. 

\begin{wrapfigure}{r}{5.8cm}\includegraphics[width=5.4cm,clip]{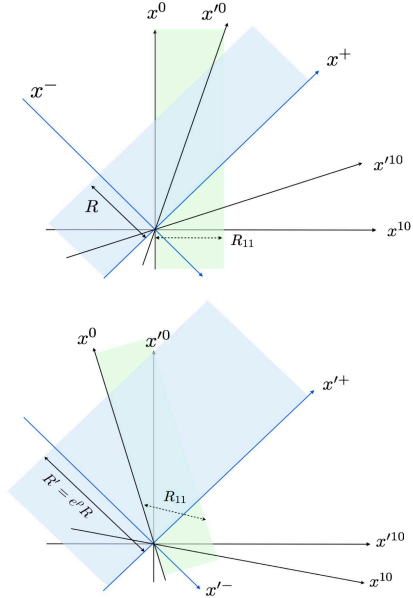}\caption{{\small DLCQ: By a boost transformation along the compactified direction $x^-$ the radius of compactification can be continuously changed. Compare the upper and lower Minkowski diagrams showing expansion of $R$ to $R'$. }}\end{wrapfigure}

Actually, it is not at all obvious whether such an 
interacting theory of particles in configuration-space formulation 
assumed in this argument 
is completely consistent, as it stands, with Lorentz invariance and 
principles of unitary quantum theory, 
within the restriction of a fixed number of particles without anti-particles. The peculiarity of 
the system such as \eqref{d0hamiltonian} is that the 
particles are interacting in a manifestly {\it non-local} fashion through 
mediating open strings, which correspond to 
off-diagonal matrix elements and 
 mix them with particle coordinates through local 
gauge transformations. 
 We can think of such mixing as an extension of 
quantum statistics of ordinary particles to 
D-branes. This is an entirely 
novel situation that we have never 
encountered previously, before the advent of 
string theory and D-brane excitations. It is 
not evident
 whether (or to what extent) our experiences with relativistic 
local field theory are applicable to this system.\footnote{It may be 
worthwhile to 
mention that this non-locality conforms to 
space-time uncertainty relation reviewed in \cite{yone1}. }  
Unfortunately, we have not acquired 
much improvement on true conceptual understandings 
on such non-locality and extended quantum statistics even 
after the two decades of various 
studies. 
 %%% Fig. 1%%%
%%%
%\begin{figure}\includegraphics[width=6cm]{dlcqboost.eps}\caption{DLCQ: By a boost transformation along the compactified direction $x^-$ the radius of compactification can be continuously changed. }\end{figure}

A very bold hypothesis made in \cite{suss}\cite{seiberg}\cite{sen}, following 
the so-called BFSS conjecture \cite{bfss} made prior to it,  is that the above 
SU($N$) gauge theory is already an exact theory 
of 11-dimensional M-theory in the special DLCQ quantization 
scheme with finite $N$. Of course, in order to exhibit 
full 11-dimensional content of this theory under this 
assumption, we should be 
able to treat continuous values of $P_{\circ}^+$ in 
any fixed Lorentz frame. Thus definitely we have to 
take the limit $N\rightarrow \infty$ and $R\rightarrow \infty$ in the end. However, 
it is quite remarkable that even a finite-$N$ theory 
may have a definite and certain exact meaning related somehow to exact non-perturbative 
formulation of M-theory. It seems a pity that 
in spite of intensive studies made from the late 90s to 
the early 2000s, progress has practically stopped 
in the last decade. 
In this lecture, I would like to revisit and 
pursue the conjecture of the DLCQ Matrix theory 
as a working hypothesis in its 
strongest form. 

For the validity of this hypothesis, there is a presumption that 
$\hat{H}$ is physically equivalent with 
the Lorentz-invariant mass-square $M^2_{{\rm eff}}$ 
for finite $N$. This must be true 
for arbitrary Lorentz transformations, which are 
not restricted to 
boosts along the compactified (10th) direction. 
Under general Lorentz 
transformation, the values of $P_{\circ}^{\pm}$ are 
mixed with transverse components $P_{\circ}^i$ of momenta. 
Therefore they must be continuously varying even with 
finitely fixed $N$. 
Here it is important to recall again that within 
the framework of the DLCQ scheme, the 
radius $R$ and hence $P_{\circ}^+$ are in fact 
regarded as continuously varying physical variables, 
since by boost transformations $x^+=x^{10}+x^0\rightarrow x'^+=
x'^{10}+x'^0=e^{-\rho}x^+, 
x^-=x^{10}-x^0\rightarrow x'^-=x'^{10}-x'^0=e^{\rho}x'^-$ along the $10$-th 
spatial direction we have transformations
\begin{align}
P_{\circ}^{\pm}\rightarrow P'^{\pm}_{\circ}=e^{\mp \rho}P_{\circ}^{\pm}
\nonumber 
\end{align}
or $R'=e^{\rho}R$ with arbitrary value of $\rho$ (see Fig. 1). 

Now the final goal of this lecture is to demonstrate 
how it is indeed possible to formulate a fully Lorentz-covariant Matrix theory such that $\hat{H}$ 
is physically equivalent to a Lorentz invariant mass-square $M_{{\rm eff}}^2$ representing the 
internal dynamics of the system. 
This will be achieved by realizing a higher gauge symmetry 
which extends the usual SU($N$) gauge symmetry, \eqref{gaugex} and \eqref{gaugea}, such that 
after imposing appropriate light-like gauge conditions  
for the higher-gauge degrees of freedom, a 
manifestly Lorentz covariant formulation which we are going to 
propose here
 reduces to the light-front Matrix theory in the physical space 
 of allowed states. 

\section{Clues toward higher-gauge symmetries}
It is obvious that, to realize such a covariant system,  
we need a new kind of symmetries which encompass and 
extend the SU($N$) gauge symmetry of the light-front 
formulation. In particular, it is crucial for the 
DLCQ scheme that such higher symmetries are operational
 even for {\it finite} $N$. In that sense, the viewpoint that 
the matrix theory is just a mere regularization of supermembranes 
should be abandoned. In fact, the simple matrix theory explained in the 
previous section exhibits  
several notable features that indeed this theory itself 
has some fundamental significance, independently of 
its relation to supermembranes. 
It is to be noted, at the basis for such features, 
that the system can be regarded as a self-consisting universal system. 
This may be signified
 in the following serial patterns of the theories with 
increasingly larger gauge groups:
\begin{center}
$\cdots\subset$ SU($N$) $\subset$ SU($N+1$)$\subset$ $\cdots \subset$ SU($N+M$) 
$\subset \cdots $, 

\vspace{0.2cm}
and 
\vspace{0.2cm}

$\cdots \subset$ SU($N_1$) $\subset$ SU($N_1$)$\times $ SU($N_2$) 
\\$\subset$ SU($N_1+N_2$) $\subset$ SU($N_1+N_2$) $\times$ SU($N_3$)
$\subset$ SU($N_1+N_2+N_3$) $\subset\cdots$, 

\vspace{0.2cm}
and so on. 
\end{center}
In other words, the system can in principle describe arbitrary multi-body 
states of physical objects which are represented by smaller sub-systems with hermitian 
sub-matrices. In this way, we can represent various many-body D-brane 
configurations and simulate their general-relativistic interactions, as reviewed in \cite{taylor}. 
For example, it has been confirmed that 3-body 
nonlinear interactions of gravitons described by 
the classical Einstein action of 11 dimensional 
supergrvatity emerge correctly\cite{oy} even with finite $N$ 
through the perturbative loop effects of off-diagonal matrix 
elements. This evidences our view that the SU($N$) matrix 
theory of finite $N$ already has some fundamental meaning 
beyond a possible
 approximate regularized formulation of supermembranes.  

With this caveat in mind, we can still extract some useful 
hints about desirable higher symmetries 
from the membrane analogy at least at a 
formal level. The SU($N$) gauge symmetry 
of light-front matrix theory corresponds mathematically to 
the {\it area}-preserving diffeomorphism \eqref{areapresdiff} 
on the membrane side. The area-preserving 
diffeomorphism can be regarded as a gauge-fixed version of 
a more general {\it volume}-preserving diffeomorphism 
represented by
\begin{align}
\delta X^{\mu}=\{F_1, F_2,X^{\mu}\}, 
\label{nambutrans}
\end{align}
which is the residual symmetry of 
 the classical action \eqref{membaction} 
after we adopt the condition $e=1$, partially 
fixing the general 3-dimensional diffeomorphism. 
One arbitrary function $F$ of the area-preserving 
transformation which corresponds to 
one arbitrary hermitian matrix $\boldsymbol{F}$ 
is now extended to two arbitrary functions 
$F_1$ and $F_2$ in \eqref{nambutrans}. 

We call the infinitesimal transformation \eqref{nambutrans} the Nambu transformation, 
since Nambu originated a generalization\cite{nambu2} of Hamiltonian 
dynamics by proposing dynamical systems in 3-dimensional 
``phase space" $(\xi_1, \xi_2,\xi_3)$ 
in which the equations of motion are 
\begin{align}
&\frac{d\xi_{a}}{dt}=\{H, G, \xi_{a}\}\equiv L\xi_{a}, 
%\frac{1}{2}\epsilon_{\ell m n}\frac{\partial (H,G)}{\partial (\xi_m, \xi_n)}
\label{nambueqmotion}
\\
&L=\sum_aL_a\frac{\partial}{\partial \xi_a}, \quad 
L_a=\sum_{b,c}\frac{1}{2}\epsilon_{abc}\frac{\partial (H,G)}{\partial (\xi_b, \xi_c)}
\label{vectorfield}
\end{align}
with two Hamitonian-like generators $H=H(\xi)$ and $G=G(\xi)$. 
In seeking for higher-gauge symmetries, it seems 
natural to try first to construct some 
matrix version of the Nambu transformation, in 
analogy with the fact that the usual SU($N$) transformation 
is the matrix version of \eqref{areapresdiff}. As a preparation 
for proceeding to such a task, I will give a brief review on some 
salient features of 
Nambu mechanics focusing its symmetry structure in the next part. 

\vspace{0.5cm}
\begin{center}
Part II : Nambu's Generalized Hamitonian Mechanics
\vspace{-0.5cm}
\end{center}
\section{A brief review of Nambu mechanics}
As a motivation for his proposal of a 
generalized Hamiltonian dynamics, Nambu stressed that  
the Liouville theorem for the Hamitonian equations of 
motion is naturally extended to \eqref{nambueqmotion} as 
\begin{align}
\sum_a\partial_aL_a=0, 
\label{liouville}
\end{align}
expressing volume-perserving nature of general 
transformation \eqref{nambutrans}. His motivation was 
a possible generalization of statistical mechanics such that 
the canonical ensemble is specified by 
two or more ``temperatures" corresponding to the 
existence of many conserved Hamiltonians. 

The usual canonical Poisson bracket is now 
replaced by a canonical ``Nambu bracket" which has 
a triplet structure:
\begin{align}
\{\xi_{a},\xi_b, \xi_c\}=1 \,\,
\mbox{iff $(a,b,c)$ is a cyclical permutation of $(1,2,3)$}. 
\label{canonambubracket}
\end{align}
A notable example from well-known dynamical systems 
which realizes this structure  
is the Euler equations of motion of a rigid top: if we regard the 3 
components $(\mu_1, \mu_2,\mu_3) $ of angular momentum 
in the body-fixed frame as canonical coordinates 
$\xi_a=\mu_a$, 
\begin{align}
\frac{d\mu_{a}}{dt}=-\sum_{b,c}\epsilon_{abc}\Bigl(\frac{1}{I_b}-\frac{1}{I_c}
\Bigr)\mu_b\mu_c=\{H, G, \mu_a\}
\label{eulereq}
\end{align}
where 
\begin{align}
H=\sum_{a}\frac{\mu_a^2}{2}, \quad 
G=\sum_{a}\frac{\mu_a^2}{2I_{a}}
\nonumber 
\end{align}
with $I_a$'s being the principal momenta of inertia of an 
assymmetrical top. 

Nambu noted that the system of equations \eqref{nambueqmotion} have 
 a ``gauge" symmetry with respect to transformation $(H, G)
\rightarrow (H', G')$ 
of the pair of Hamiltonians defined by $\frac{\partial(H',G')}{\partial(H,G)}=1$ which can be expressed equivalently as
\begin{align}
H\delta G-H'\delta G'=\delta f \quad 
or \quad \frac{\partial f}{\partial G}=H, \,\, 
\frac{\partial f}{\partial G'}=-H', 
\label{ngaugetrans}
\end{align}
where $f=f(G,G')$ is an arbitrary function of $G$ and $G'$ 
as a generating function which implicitly 
determines the transformation. He also correctly noted that this is not the most general 
gauge transformation from the viewpoint of 
general volume-preserving transformations. 
In the latter viewpoint, \eqref{ngaugetrans} would be 
slightly generalized to the transformations of the following form:
\begin{align}
H\frac{\partial G}{\partial \xi_a}-H'\frac{\partial G'}{\partial 
\xi_a}=\frac{\partial S}{\partial \xi_a}, 
\label{generalgauge}
\end{align}
with an arbitrary function $S=S(\xi_a)$ of $\xi_a$'s, instead of 
the form $f(G(\xi),G'(\xi))$. The reason for Nambu's remark is 
originated from the fact that 
$L_a$ defined in \eqref{vectorfield} is not 
the most general form satisfying \eqref{liouville}, even if 
we consider arbitrary pair $(H,G)$.  
Locally, the most general form of the vector $L_a$ is 
\begin{align}
L_a=\frac{1}{2}\sum_{b,c}\epsilon^{abc}(\partial_bA_c-\partial_cA_a)
\label{generalform}
\end{align}
with some vector gauge field $A_a(\xi)$, in terms of which the 
general gauge transformation keeping 
$L_a$ invariant is  
\begin{align}
\delta A_a=\partial_aS. 
\label{puregauge}
\end{align}
This would lead to \eqref{generalgauge}. 
The form \eqref{vectorfield} corresponds to a special form
\begin{align}
A_a=H\partial_aG.
\nonumber 
\end{align}
However, the form \eqref{puregauge} is not ensured in general by 
\eqref{ngaugetrans} for an arbitrary scalar function $S(\xi)$.\footnote{If we suppose that the general form 
of the gauge transformation could be
 realized in the form \eqref{generalgauge}, 
it should be possible to adopt, say, the ``axial gauge" $\partial_3G'=0$, which however requires that $\{H,G,\xi_1\}_{{\rm N}}/\{H,G,\xi_2\}_{{\rm N}}=\{H',G',\xi_1\}_{{\rm N}}/\{H',G',\xi_2\}_{{\rm N}}=-\partial_2G'/\partial_1G'$ is independent 
of $\xi_3$. 
This is impossible for most general choice of $(H,G)$. 
Whether this is possible thus depends on a particular 
system we consider. Incidentally, the case of the 
Euler equation \eqref{eulereq} is a typical example where this gauge choice 
is allowed. } 
The situation is in contrast to ordinary Hamilton 
mechanics where the vector field $L_a=\sum_b\epsilon^{ab}
\partial_bF$ with an arbitrary scalar field $F$ locally exhausts  
the area-preserving condition $\partial_aL_a=0$.    
Nambu suggested that, to 
exhaust the most general form satisfying the latter condition in terms of triple bracket, the equations of motion 
may be extended to 
\begin{align}
\frac{d\xi_{\ell}}{dt}=\sum_i\{H_i,G_i, \xi_{\ell}\}, \quad 
or \quad A_a=\sum_iH_i\partial_aG_i
\label{severalHG}
\end{align}
by introducing multiple pairs $(H_i, G_i)$ instead of a single 
pair $(H, G)$. Then in general there is no manifestly 
conserved ``Hamiltonians", somewhat contrary to 
Nambu's original motivation for extending 
statistical mechanics. With this generalization, the above gauge 
transformation must be generalized to
\[
\sum_i(H_i\delta G_i-H'_i\delta G'_i)=\delta f, 
\quad 
\frac{\partial f}{\partial G_i}=H_i,\,\, 
\frac{\partial f}{\partial G'_i}=-H'_i. 
\]
which then allows one to 
set
\[
\sum_i\Bigl(H_i\frac{\partial G_i}{\partial \xi_a}-H'\frac{\partial G'_i}{\partial \xi_a}\Bigr)=\frac{\partial S}{\partial \xi_a}
\]
for an arbitrary function $S(\xi)$. 

If we had started from the 
general form \eqref{generalform} for the equations of motion 
from the beginning, 
a motivation for introducing the 
triple bracket and canonical structure such as \eqref{canonambubracket} 
would not arise, since then the role of 
Hamiltonians would have been played 
directly by the vector gauge fields $A_a$: no need to introduce 
pairs 
$(H_i, G_i)$. 
In that sense, it was fortunate for us that Nambu 
insisted on generalizing Hamilton mechanics 
in his way using the triple bracket.

Nambu further studied canonical transformations  
$\xi_a\rightarrow \xi'_a$ of canonical coordinates which preserve \eqref{canonambubracket}. 
Restricting to the simplest case of linear transformations, 
he noticed disappointedly that there is a difficulty in extending the canonical structure to higher-dimensional phase space, 
 $\{(\xi^{(p)}_1,\xi_2^{(p)},\xi_3^{(p)}); p=1,\ldots, n\}$,  of $3n$-dimensions, 
  on the basis of a naturally looking postulate that the canonical 
bracket obeys 
\begin{align}
\{\xi_{\ell}^{(p)}, \xi_m^{(q)}, \xi_n^{(r)}\}
\equiv 1 \,\, &\mbox{iff } (\ell, m,n)\in \mbox{cyclic permutations of } (1,2,3)\nonumber \\
&\mbox{ and } p=q=r, 
\label{manybody}
\end{align}
in a naive analogy with the Hamilton mechanics. 
The problem is that the equations \eqref{nambueqmotion} 
cannot 
preserve this canonical structure whenever the time 
development mixes different triplets with different $p$'s. 
This implies that from the viewpoint of canonical 
structure it is not possible to extend 
the Nambu mechanics to coupled many-body systems, in spite 
of several subsequent attempts toward such directions.\footnote{
Nambu himself alluded to a model 
which simulates coupled spin systems. However, that has never 
been published, unfortunately.
} 

On the other hand, it is easy to generalize this system to an  
$N$-dimensional phase space $(\xi_1,\ldots, \xi_N)$ such that 
the time evolution is described by a set  
of $N-1$ Hamitonians $(H_1,H_2,\ldots, H_{N-1})$  :
\begin{align}
\frac{d\xi_a}{dt}=
\{H_1,H_2,\ldots, H_{
N-1},\xi_a\}
\equiv \frac{\partial(H_1, \ldots,H_{N-1}, \xi_a)}{
\partial(\xi_1,\ldots, \xi_{N-1},\xi_N)}. 
\nonumber 
\end{align}
Obviously, this preserves the $N$-dimensional volume as 
a straightforward extension of the case $N=3$. 
With respect to symmetries, these extended systems inherit  
the same problems as in $N=3$ with respect to possible extensions to 
$nN$-dimensional phase space. 
In the present lecture, we restrict ourselves only to the case $N=3$ and 
$n=1$. 

One of Nambu's further concerns was 
to examine whether or not the above structure could be 
extended to quantum theory. For that purpose, he considered 
the problem how the triple bracket defined by 
Jacobian determinant in classical theory can be 
mapped to some algebraic structure denoted 
by $[A,B,C]$, which is required to preserve  
the basic properties of the classical bracket, namely, 
\begin{enumerate}
\item[(a)] skew symmetry:
\begin{align}
[A,B,C]=-[B,A,C]=[B,C,A]=\cdots, 
\label{skew}
\end{align}
\item[(b)] derivation law:
\begin{align}
[A_1A_2, B,C]=[A_1, B, C]A_2+A_1[A_2,B,C], 
\quad {\rm etc}.
\label{derivation}
\end{align}
\end{enumerate}
In particular, he postulates the following triple commutator 
as a candidate for quantum triple bracket:
\begin{align}
[A,B,C]_{{\rm N}}&\equiv ABC+BCA+CAB-BAC-ACB-CBA\nonumber \\
&=A[B,C]+B[C,A]+C[A,B], 
\label{triplecommu}
\end{align}
and correspondingly the generalized 
Heisenberg equations of motion, 
\begin{align}
i\frac{dF}{dt}=[H,G,F]_{{\rm N}}. 
\nonumber 
\end{align}
In this definition, only the property (a) is manifestly 
satisfied, but not (b) automatically. So he discussed  
various possibilities of algebraic structures for the set of 
operators $H, G, \ldots, F, \ldots $, 
including possible generalizations as 
\eqref{severalHG}, by studying slightly weakened versions of 
these conditions. 
Unfortunately, however, the main conclusion\footnote{To quote his own words, ``{\it One is repeatedly led to discover that the 
quantized version is essentially equivalent to the ordinary 
quantum theory. This may be an indication that quantum 
theory is pretty much unique, although its 
classical analogue may not be.}"} was that 
it was difficult to realize quantization nontrivially. 
Nambu then studied the possibilities of using non-associative 
algebras, but his conclusion was again not definitive.

\section{The fundamental identity and 
canonical structure}
Further developments of Nambu mechanics rested largely upon 
a seminal work by Takhtajan\cite{takh} which appeared 
after two decades since Nambu's original proposal. 
In this work, it was pointed out that there exists an 
important identity (now known as the ``Fundamental 
Identity", FI) satisfied by the Nambu bracket, which generalizes the Jacobi 
identity in the case of Poisson bracket. 
For an arbitrary set of five functions $(F_1,\ldots, F_5)$, 
it takes the form
\begin{align}
&\{F_1,F_2,\{F_3,F_4,F_5\}\}=\{\{F_1,F_2,F_3\},F_4,F_5\}
\nonumber \\
&+\{F_3, \{F_1,F_2,F_4\},F_5\}+\{F_3,F_4,\{F_1,F_2,F_5\}\}.
\label{FI}
\end{align}
This ensures that the canonical structure \eqref{canonambubracket} 
is preserved by the time evolution described by \eqref{nambueqmotion}, as can be seen by applying 
this identity with $F_1=H, F_2=G$ and $F_3=\xi_1,F_4=\xi_2, 
F_5=\xi_3$. The same can be said for general infinitesimal 
canonical transformation defined by
\begin{align}
\delta \xi_a=\{F_1,F_2,\xi_a\}
\label{genecano}
\end{align}
with a pair $(F_1,F_2)$ of arbitrary functions of 
the canonical coordinates. 
This clarifies the reason why it is difficult to 
generalize the system to interacting many-body cases. 
For instance, the postulate \eqref{manybody} does not in fact 
satisfy the FI. This is in marked contrast to ordinary 
Hamiltonian mechanics. Also, 
the triple commutator \eqref{triplecommu} 
does not in general satisfy the FI. This partially explains the 
difficulties encountered in quantization. 

These features indicate that the Nambu mechanics 
is a quite restricted dynamical system which is characterized 
by the stringent structure of the FI. In other words, 
we cannot expect the same kind of universality for Nambu mechanics as we have in 
the framework of Hamilton mechanics. 

Nevertheless, 
we may also take a viewpoint, which is complementary 
to the foregoing statement,  that Nambu mechanics 
provides a new structure characterized by higher-symmetry transformations \eqref{genecano} 
with two arbitrary functions as parameters of transformation, 
instead of corresponding transformation with one arbitrary 
function in Hamilton mechanics. 
Our standpoint in applying and extending
 the Nambu transformation starting from \eqref{nambutrans} is 
 this interpretation of Nambu mechanics. 
Instead of developing further the Nambu mechanics as 
a dynamical system, 
we extract only new symmetry structure as a clue 
toward higher symmetries which we need for 
constructing a covariant version of Matrix theory. 
It is possible to imagine dynamical systems  
which obey the usual Hamiltonian mechanics with respect to 
its time evolution, 
but equipped with higher symmetries characterized by 
some appropriate (quantized or discretized) version of 
\eqref{nambutrans} which enables us to encompass 
the usual SU($N$) transformation \eqref{gaugex} and 
\eqref{gaugea} as a special (gauge-fixed) case. 
This is precisely what we are going to try in the third part of the 
present lecture. 

\section{Further remarks on the nature of Nambu mechanics}
Before proceeding to exploration toward such a direction, 
I would like to make 
further comments on the nature of Nambu mechanics. One important remark made in 
\cite{takh} is that we can always regard the Nambu mechanics 
as a special case of Hamiltonian mechanics. 
Namely, given the structure satisfying the FI, 
we can always define Poisson brackets which are subordinated to 
Nambu bracket, by
\begin{align}
\{F, G\}_H=-\{F, G\}_H\equiv 
\{F,H,G\}
\nonumber 
\end{align}
where $H$ is arbitrary but fixed once and for all. 
It is easy to see that by setting $F_2=F_4=H$, 
the FI \eqref{FI} reduces to the Jacobi 
identity for this Poisson bracket
\begin{align}
\{F_1,\{F_3,F_5\}_H\}_H=\{\{F_1,F_3\}_H,F_5\}_H+
\{F_3,\{F_1,F_5\}_H\}_H, 
\nonumber 
\end{align}
and the Nambu equations of motion 
now take the form,
\begin{align}
\frac{dF}{dt}=-\{G, F\}_H
\end{align}
with a single Hamiltonian $G$. Another Hamiltonian 
$H$ now characterizes the structure of phase space 
through the Poisson bracket. 
This fact strengthens our view that the usual 
Hamilton mechanics is far more universal as a scheme for 
representing dynamics, 
and Nambu mechanics should be regarded as a 
special case of it characterized by higher symmetries, 
rather than as a new universal framework for representing 
dynamics. 
In fact the Euler equations can also be formulated in terms of the standard 
Poisson brackets of the angular momenta of the body-fixed frame 
which are in fact nothing 
but this representation: namely we have 
\[
\{\mu_a, \mu_b\}_H=-\sum_c\epsilon_{abc}\mu_c. 
\]

Thus it is not unreasonable to 
take the standpoint, for arbitrary Nambu system 
of equations of motion, that quantization as a means 
to develop a new dynamical approach should 
be done by elevating the subordinated Poisson 
brackets to commutators in an appropriate Hilbert 
space which provides a representation of the 
commutator algebra corresponding to a 
chosen 
subordinate Poisson bracket. 
This may not be along Nambu's original 
intension, but certainly is a possible 
and consistent attitude. 
In view of the presence of gauge symmetry \eqref{ngaugetrans} 
which is intrinsic to the Nambu system, one of 
main issues from this viewpoint would then 
be whether or not this provides physically unique 
result for different but gauge-equivalent choices of 
$(H,G)$, rather than trying to quantize Nambu brackets directly. 
 The simplest case is just an 
interchange of $H$ and $G$ or $(H,G)
\rightarrow (G,-H)$ corresponding to the generating 
function 
$f=HG$. We can also 
arbitrary mix these two 
Hamiltonians. In other words, we need to extend the framework of 
quantum mechanics such that these gauge transformations 
as well as the canonical transformation of coordinates $\xi_a$ 
can act in a covariant fashion in the space of 
physical states. 
In the case of the Euler top, for example, 
it seems that the situation is quite non-trivial from 
this viewpoint. 
This question reminds us of Nambu's remark quoted 
in the footnote in the end of the previous section, though 
of course in 
a different context. To the author's knowledge, there is 
practically no work done from this standpoint yet. 

Another issue closely related to the above question of 
quantized Nambu mechanics 
is the Hamilton-Jacobi theory of Nambu mechanics. 
The latter would be a possible clue toward quantization, 
remembering Schr\"{o}dinger's approach to quantum mechanics. 
This problem also seems not to be discussed seriously. In existing literature, the problem of quantization 
has been discussed mostly from the algebraic point of view 
of realizing Nambu bracket in some operatorial or matrix form. 
Possible ``wave-mechanical" approaches are quite scarce, if not none 
when we include passing expectations or remarks such as, say, a path-integral approach as already mentioned in \cite{takh}. It seems fair to 
say that such possibilities have never been pursued to 
appropriate depth. 
%One of possible starting points towards such a direction has been laid down, at least partially,  in this last referencein which an action principle for Nambu mechanics was also given. 
These problems will be discussed in separate 
publications,\cite{yone-toapp} since they are 
rather remote from our present purpose 
of pursuing a covariant Matrix theory.% now in preparation\cite{yoneHJ}. 

\vspace{0.7cm}
\begin{center}
Part III: Higher Symmetries and Covariantized Matrix Theory
\end{center}
\vspace{-0.4cm}
\section{A matrix version of Nambu bracket and higher 
symmetry}
Now we come to our main subject of this lecture. 
We will essentially follow the paper\cite{yone2} to which I would like to refer readers for more detail, 
including references. 
As explained in section 4, the Nambu transformation 
\eqref{nambutrans} 
\[
\delta X^{\mu}(\xi)=\{F_1, F_2, X^{\mu}(\xi)\}
\]
can be a starting point for exploring possible 
higher-symmetries which generalize the usual 
SU($N$) transformation of the light-front 
Matrix theory. For that purpose, it is 
necessary to find an appropriate 
counterpart of the Nambu bracket in 
matrix algebra. 
We have initiated such a project of  
quantizing or more appropriately discretizing 
the Nambu bracket in ref.\cite{almy}. 
Unfortunately, however, 
 we could not present 
appropriate application of our work to 
construct covariant Matrix theory 
at that time. 
One of our proposals for realizing 
the FI using discretized algebraic structures 
was 
\begin{align}
[A, B, C]_{{\rm almy}}=
({\rm Tr} \,\boldsymbol{A})
[\boldsymbol{B},\boldsymbol{C}]+({\rm Tr}\, \boldsymbol{B})[\boldsymbol{C},\boldsymbol{A}]+({\rm Tr}\, \boldsymbol{C})
[\boldsymbol{A},\boldsymbol{B}], 
\label{almy}
\end{align}
using ordinary $N\times N$ hermitian matrices. 
For the validity of the FI, actually, the use of 
the matrix traces ${\rm Tr}\,A$ etc is {\it not} 
essential. We can replace them by any 
single auxiliary but independently 
varying (real) numbers, denoted by $A_{{\rm M}}, B_{{\rm M}}$ etc,  associated {\it separately} with each matrix variable, namely
\begin{align}
[A,B,C]=A_{{\rm M}}[\boldsymbol{B},\boldsymbol{C}]+B_{{\rm M}}[\boldsymbol{C},\boldsymbol{A}]+C_{{\rm M}}
[\boldsymbol{A},\boldsymbol{B}], 
\nonumber 
\end{align}
which we will adopt exclusively in the following. Unnecessary identification 
of the auxiliary variable with trace was 
one of stumbling blocks by which we were stuck in our original work. 

Note that this form,  as well as the above original form using trace, is automatically skew symmetric. 
On the other hand, {\it neither} does satisfy the derivation property 
(Nambu's criterion (b), \eqref{derivation}) 
for general matrix products. Although this might look as 
a deficiency from the viewpoint of constructing a 
universal framework of Nambu mechanics, our standpoint 
is different as we have already discussed emphatically 
in Part II. From the viewpoint of 
symmetry, this deficiency rather turns to a merit in that 
it means stronger constraints in 
constructing theory than the case with automatic presence of 
derivation property. 

In \cite{almy}, we have also proposed alternative directions in which 
the matrices are replaced by ``cubic matrices" $A_{abc}, B_{abc}, \ldots, $ 
with three indices. An example is
\[
[A,B,C]_{{\rm cubic}}=(ABC)+(BCA)+(CAB)-(CBA)-(ACB)-(BAC)
\] 
where 
\[
(ABC)_{abc}=\sum_{p}A_{abp}\langle B\rangle C_{pbc}
\equiv\sum_{pqm}A_{abp}B_{qmq}C_{pbc}.
\]
These and similar possibilities might still be useful in 
different context: for instance, we may try to 
regularize the covariant action of supermembrane, directly, 
without relying on the DLCQ interpretation, following 
the original and primitive motivation from which we have 
started to explain matrix theories. 
 In the following, however, we do not 
pursue such possibilities. 

It should be noted that the object $[A,B,C]$ itself 
can be treated as a (anti-hermitian) matrix; we {\it define} the would-be auxiliary 
variable associated with this matrix is zero:
\[
[A,B,C]_{{\rm M}}\equiv 0.
\]
Our original form \eqref{almy} using trace is just a special case where this is automatically 
satisfied without demanding it explicitly. 
By a straightforward calculation, it is easy to 
confirm that the FI is valid:
\begin{align}
[A,B,[C,D,E]]=[[A,B,C],D,E]+[C,[A,B,D],E]+[C,D,[A,B,E]].
\nonumber 
\end{align}
A crux of such a calculation is that the terms involving the 
commutator $[\boldsymbol{A}, \boldsymbol{B}]$ cancel 
among themselves on the r.h.side, to be 
consistent with the l.h.side with $[C,D,E]_{{\rm M}}\equiv 0$. 
The remaining terms are arranged into 
the form coinciding with the l.h.side using 
the ordinary Jacobi identities for matrix commutators. 

Now, the dynamical variables and also the 
parameters of higher transformations are in general 
a set of matrices and associated auxiliary 
variables which are denoted by $A=(A_M, \boldsymbol{A}), \ldots, etc $. 
Thus we denote the space-time coordinate variables 
by $X^{\mu}(\tau)=(X^{\mu}_{{\rm M}}(\tau), \boldsymbol{X}^{\mu}(\tau))$. Here we have introduced a Lorentz invariant (proper) time 
parameter $\tau$.  The roles of $\tau$ and of the 
auxiliary variables $X^{\mu}_{{\rm M}}(\tau)$ will be
discussed later. 

The higher transformations are then defined to be 
\begin{align}
\delta X^{\mu}=i[F,G,X^{\mu}]
\nonumber 
\end{align}
with two ``parameters", $F=(F_{{\rm M}}, \boldsymbol{F})$ and 
$G=(G_{{\rm M}}, \boldsymbol{G})$ of {\it local} transformations, both of 
which are arbitrary functions of time. Therefore 
the auxiliary variable of these spacetime 
coordinate variables are invariant under higher transformations by definition, 
\begin{align}
\delta X^{\mu}_{{\rm M}}=0, 
\nonumber 
\end{align}
while their matrix part is transformed as
\begin{align}
\delta \boldsymbol{X}^{\mu}=i[F_{{\rm M}}\boldsymbol{G}
-G_{{\rm M}}\boldsymbol{F}, \boldsymbol{X}^{\mu}]+
i[\boldsymbol{F},\boldsymbol{G}]X_{{\rm M}}^{\mu}. 
\nonumber 
\end{align}
The first term takes the form of usual 
SU($N$) (infinitesimal) unitary transformation with the 
hermitian matrix $F_{{\rm M}}\boldsymbol{G}
-G_{{\rm M}}\boldsymbol{F}$. The second term represents a 
shift of the matrix. Due to this term, we can shift $\boldsymbol{X}
^{\mu}$ using the 
traceless matrix $i [\boldsymbol{F},\boldsymbol{G}]$ which is 
almost (but not completely) independent of the first term.  
As in the case of the Nambu equations of motion, 
we can treat this shift as being completely independent 
of the first term by a slight generalization. Namely, in analogy with 
\eqref{severalHG}, we generalize the transformation 
by introducing an arbitrary number of pairs 
$(F^{(r)}, G^{(r)})$ instead of a single pair $(F,G)$ 
to 
\begin{align}
\delta_{HL}X^{\mu}=\delta_HX^{\mu}+\delta_LX^{\mu}=
(0, i[\boldsymbol{H},\boldsymbol{X}^{\mu}])+(0, \boldsymbol{L}X_{{\rm M}}^{\mu})
\label{HLtransX}
\end{align}
where
\begin{align}
\boldsymbol{H}&\equiv \sum_r F_{{\rm M}}^{(r)}\boldsymbol{G}^{(r)}-G_{{\rm M}}^{(r)}\boldsymbol{F}^{(r)}, \nonumber  \\
\boldsymbol{L}&\equiv i\sum_r[\boldsymbol{F}^{(r)}, \boldsymbol{G}^{(r)}]\nonumber 
\end{align}
are now regarded as two independent (traceless) hermitian matrices. 
In this form, there is no problem associated with ``gauge" symmetry \eqref{generalgauge} 
in the sense worried by Nambu. 
Of course, once we have this form, we could 
actually forget about its origin from Nambu bracket. 
Our standpoint would coincide with my previous remark 
on the direct use of vector gauge field $A_a$ in section 5, concerning 
the meaning of the general form \eqref{generalform} in Nambu mechanics. 
Even if so, however, the bracket notation will still be very 
useful and convenient in expressing invariants succinctly. 

The shift term enables one to eliminate the 
traceless part of any single matrix, whenever 
the auxiliary variable associated with it is not zero, by a local 
gauge transformation. For example, 
if $X^0_{{\rm M}}$ is non-zero, we can 
transform the martrix $\boldsymbol{X}^0$ into the 
unit matrix up to a single proportional function.
% as\[\boldsymbol{X}^0(\tau)\rightarrow X^0(\tau)\begin{pmatrix} 1 & 0 & 0 &. & \ldots \ldots \ldots&0 \\ 0 & 1 & 0&. & \ldots \ldots \ldots&0\\ 0 & 0 & 1 &. &\ldots\ldots \ldots& 0\\   . &    &    &. & \ldots \ldots\ldots&0 \\. &    &    &. & \ldots \ldots\ldots&0\\. &    &    &. & \ldots \ldots\ldots&0\\0 & 0 & 0& .& \ldots \ldots\ldots& 1\end{pmatrix}.\]

Now the next important question is this: what are, if any, invariants  
under these higher transformations? Obviously, 
usual traces of matrix products, such as ${\rm Tr}(\boldsymbol{X}\boldsymbol{Y})$,  
cannot in general be invariant, unless $X_{{\rm M}}=0=Y_{{\rm M}}$ 
which however seems to render the higher part of the transformations 
ineffective. There is a simple resolution. The matrices should appear only 
through
 triple brackets, for which themselves
  the auxiliary M-components are equal to 
zero by definition. The simplest non-trivial example is, 
with arbitrary two sets of variables $(A,B,C)$ and $(X,Y,Z)$,  
\begin{align}
\langle [A,B,C], [X,Y,Z]\rangle& \equiv 
{\rm Tr} \Bigl((A_{{\rm M}}[\boldsymbol{B},\boldsymbol{C}]
+B_{{\rm M}}[\boldsymbol{C},\boldsymbol{A}]
+C_{{\rm M}}[\boldsymbol{A},\boldsymbol{B}])\nonumber \\
&\times (X_{{\rm M}}[\boldsymbol{Y},\boldsymbol{Z}]
+Y_{{\rm M}}[\boldsymbol{Z},\boldsymbol{X}]
+Z_{{\rm M}}[\boldsymbol{X},\boldsymbol{Y}])\Bigr).
\nonumber 
\end{align}
Because the FI is valid for each component $r$, 
\[
[F^r,G^r,[A,B,C]]=[[F^r,G^r,A],B,C]+[A,[F^r,G^r,B],C]
+[A, B,[F^r,G^r,C]], 
\]
it is valid after summing over $r$ too. This means that the 
derivation (or distribution) law 
\[
\delta_{HL}[A,B,C]=[\delta_{HL}A, B, C]+
[A, \delta_{HL}B, C]+[A, B, \delta_{HL}C]
\]
is valid inside the bracket with 
respect to our higher gauge transformation. Therefore 
we have, remembering $[A,B,C]_{{\rm M}}=[X,Y,Z]_{{\rm M}}=0$, 
\[
\delta_{HL}[A,B,C]=i[\boldsymbol{H}, [A,B,C]], 
\quad 
\delta_{HL}[X,Y,Z]=i[\boldsymbol{H}, [X,Y,Z]], 
\]
which ensures 
\begin{align}
\delta_{HL}\langle [A,B,C], [X,Y,Z]\rangle=&\langle \delta_{HL}[A,B,C], [X,Y,Z]\rangle+\langle [A,B,C], \delta_{HL}[X,Y,Z]\rangle\nonumber \\
=&0. 
\nonumber 
\end{align}

This result indicates that, corresponding to the 
potential term ${\rm Tr}[\boldsymbol{X}_i,\boldsymbol{X}_j]^2$ in the light-front Matrix theory, we have a simple integral 
invariant composed of the coordinate matrices
\begin{align}
&\frac{1}{12}\int d\tau \,e\, \langle [X^{\mu},X^{\nu},X^{\sigma}]
[X_{\mu},X_{\nu},X_{\sigma}]\rangle \nonumber \\
=&\frac{1}{4}\int d\tau \, e\, {\rm Tr}
\Big(X_{{\rm M}}^2
[\boldsymbol{X}^{\nu},\boldsymbol{X}^{\sigma}][\boldsymbol{X}_{\nu},\boldsymbol{X}_{\sigma}]
-2[X_{{\rm M}}\cdot \boldsymbol{X}, \boldsymbol{X}^{\nu}]
[X_{{\rm M}}\cdot\boldsymbol{X}, \boldsymbol{X}_{\nu}]
\Bigr)
\label{potential}
\end{align}
where by ($\cdot$) we denote the usual Lorentz invariant scalar 
product, and the symbol $e=e(\tau)$ is the ein-bein, transforming as 
a density ($e(\tau)d\tau=e'(\tau')d\tau'$) under arbitrary reparametrization of the time parameter 
$\tau$. Clearly, the above form of the 
potential term is contained in the first term of this expression, 
if we are allowed to identify the Lorentz invariant $X_{{\rm M}}^2$ with the M-theory 
parameters appropriately. 
Later we will examine this question and also whether other terms may be 
ignored in the physical space. 

\section{Lorentz-invariant canonical formalism of higher symmetries 
with further extensions}
We treat this dynamical system by a canonical formalism 
with respect to a single Lorentz-invariant time parameter $\tau$, and  introduce momentum variables, denoted by $(P_{{\rm M}}^{\mu}, \boldsymbol{P}^{\mu})$, which are canonically conjugate in the usual sense to the 
coordinate variables $X^{\mu}=(X_{{\rm M}}^{\mu}, \boldsymbol{X}^{\mu})$. The canonical Poisson brackets are thus 
\begin{align}
\{X^{\mu}_{{\rm M}}, P_{{\rm M}}^{\nu}\}_{{\rm P}}&=
\eta^{\mu\nu}, \nonumber \\
\{ X^{\mu}_{ab}, P^{\nu}_{cd}\}_{{\rm P}}
&=\delta_{ad}\delta_{bc}\eta^{\mu\nu}, \nonumber 
\end{align}
with all other Poisson brackets being zero 
({\it e.g.} $\{X_{ab}^{\mu}, P_{{\rm M}}^{\nu}\}_{{\rm P}}=0$, etc). 
Note that the appearance of the indefinite 11-dimensional 
Minkowkian metric $\eta^{\mu\nu}$ is due to our fundamental 
requirement of 11-dimensional Lorentz covariance. 

Perhaps, some of you may wonder about the feasibility 
of only a single proper time, in spite of the fact 
that we are here dealing with a many-particle theory. 
In a standard method of treating many-particles 
relativistically, we usually introduce proper time
for each particle separately. 
In our case, however, 
that is very difficult to do, since we cannot actually 
separate particle degrees of freedom and the other 
degrees of freedom which mediate interactions among them. 
This peculiarity has been already emphasized in Part I of this lecture. 
It is more natural to describe the dynamics 
using a single global (but Lorentz invariant) ``time" synchronized independently 
of the sizes of matrices to all subsystems, 
when we decompose a system into several subsystems, 
since they are interacting non-locally, once we adopt 
the description of Matrix theory.

We demand that the canonical brackets are invariant under 
the higher transformations. 
%In other words, what we are going to do is to construct a symplectic structure in a way which is invariant under Lorentz transformations and higher gauge symmetries. 
This requirement fixes the 
transformation laws of 
the momentum variables as
\begin{align}
\delta_{HL}\boldsymbol{P}^{\mu}&=i[\boldsymbol{H},
\boldsymbol{P}^{\mu}]\equiv\delta_H\boldsymbol{P}^{\mu}, 
\quad 
\delta_{HL} P_{{\rm M}}^{\mu}=
-{\rm Tr}\Bigl(%[\boldsymbol{F}, \boldsymbol{G}]
\boldsymbol{L}\boldsymbol{P}^{\mu}\Bigr)\equiv\delta_LP_{{\rm M}}^{\mu}.
\label{HLtransP}
\end{align}
The generator of the higher transformations with respect 
to the Poisson brackets is 
\begin{align}
{\cal C}_{HL}\equiv {\rm Tr}
\Bigl(
\boldsymbol{P}_{\mu}\bigl(i[\boldsymbol{H}, \boldsymbol{X}^{\mu}]+\boldsymbol{L}X_{{\rm M}}^{\mu}
\bigr)\Bigr), 
\label{canogene}
\end{align}
by which the transformation of an arbitrary 
functions $O=O( X_{{\rm M}}, \boldsymbol{X}, 
P_{{\rm M}}, \boldsymbol{P})$ takes the form $
\delta_{HL}O=\{O, {\cal C}_{HL}\}
$.
Since the transformation $\delta_{HL}\boldsymbol{P}^{\mu}\equiv \delta_H\boldsymbol{P}^{\mu}$ coincides with the ordinary SU($N$) transformation, we have an integral invariant, simply by taking 
the trace of any product of momentum matrices, as
\begin{align}
\int d\tau \, e\, {\rm Tr}(\boldsymbol{P}\cdot \boldsymbol{P}).
\label{momentum2}
\end{align}
This is in contrast to the coordinate matrices, 
where there is a shift term in $\delta_{HL}\boldsymbol{X}^{\mu}$ but no transformation of the M-variable. In the case of momentum, 
the M-variable has a shift-type transformation instead of the matrix variables. Thus the 
usual kinetic term is not allowed  for $P_{{\rm M}}^{\mu}$ as it stands. 

Together with the integral invariant 
corresponding to the potential term,  it is important to notice that our system has 
a simple global symmetry under scaling $\tau 
\rightarrow \lambda^2\tau$ of the propertime 
parameter:
\begin{align}
&\boldsymbol{X}^{\mu}\rightarrow \lambda \boldsymbol{X}^{\mu}, 
\quad 
X_{{\rm M}}^{\mu}\rightarrow \lambda^{-3}X_{{\rm M}}^{\mu},\quad \boldsymbol{P}^{\mu}\rightarrow \lambda^{-1}\boldsymbol{P}^{\mu}, 
\quad 
P_{{\rm M}}^{\mu}\rightarrow \lambda^{-3}P_{{\rm M}}^{\mu},
\label{scaling}
\end{align}
Later we will argue that this scale symmetry will govern the fundamental scales of this theory 
as is expected to be a possible non-perturbative formulation 
of M-theory. Note that 
we here assume that the ein-bein $e(\tau)$ transforms as a 
{\it dimensionless} 
scalar, $e\rightarrow e$, under this global scale transformation unlike the 
case of {\it local} reparametrizations.

Now since the higher transformations are local with respect to 
$\tau$, we have to use covariant derivatives by introducing 
gauge fields in order to properly deal with their evolutions in 
$\tau$. We need two independent 
matrix gauge fields corresponding to $H$ and $L$ transformations, 
denoted by $\boldsymbol{A}$ and $\boldsymbol{B}$, 
respectively. Both are by definition 
{\it traceless} $N\times N$ hermitian matrices. 
Their transformation laws are 
\begin{align}
\delta_{HL}\boldsymbol{A}&=i[\boldsymbol{H},\boldsymbol{A}]-\frac{1}{e}\frac{d}{d\tau}\boldsymbol{H}\equiv -\frac{1}{e}
\frac{D\boldsymbol{H}}{D\tau}, \nonumber \\
\delta_{HL}\boldsymbol{B}&=i[\boldsymbol{H}, \boldsymbol{B}]-i[\boldsymbol{L},\boldsymbol{A}]+\frac{1}{e}\frac{d}{d\tau}\boldsymbol{L}
\equiv i[\boldsymbol{H},\boldsymbol{B}]+
\frac{1}{e}\frac{D\boldsymbol{L}}{D\tau}, \nonumber 
\end{align}
where the gauge fields are defined to be 
scalars under $\tau$-reparametrization, as signified 
by the presence of the ein-bein associated with the 
time differential. 
Note that we do not assign auxiliary variables for 
the gauge-field matrices, and also that 
the scaling of gauge fields and that of the parameters of 
transformations are
\[
\boldsymbol{A}\rightarrow \lambda^{-2}\boldsymbol{A}, 
\quad 
\boldsymbol{B}\rightarrow \lambda^2\boldsymbol{B}, 
\quad \boldsymbol{H}\rightarrow \boldsymbol{H}, 
\quad 
\boldsymbol{L}\rightarrow \lambda^4\boldsymbol{L}. 
\]

The covariant derivatives of the coordinate and momentum 
variables are then given by
\begin{align}
\frac{D'X^{\mu}}{D\tau}&=\Bigl(\frac{dX_{{\rm M}}^{\mu}}{d\tau}, 
\frac{D'\boldsymbol{X}^{\mu}}{D\tau}\Bigr), 
\quad \frac{D'P^{\mu}}{D\tau}=\Bigl(\frac{dP_{{\rm M}}^{\mu}}{d\tau}, 
\frac{D'\boldsymbol{P}^{\mu}}{D\tau}\Bigr), \nonumber \\
\frac{D'\boldsymbol{X}^{\mu}}{D\tau}&=
\frac{d\boldsymbol{X}^{\mu}}{d\tau}+ie[\boldsymbol{A}, 
\boldsymbol{X}^{\mu}]-e\boldsymbol{B}X^{\mu}_{{\rm M}},\nonumber \\
\frac{D'\boldsymbol{P}^{\mu}}{D\tau}
&= \frac{d\boldsymbol{P}^{\mu}}{d\tau}+
ie[\boldsymbol{A}, \boldsymbol{P}^{\mu}], 
\nonumber \\
\frac{D'P_{{\rm M}}^{\mu}}{D\tau}&\equiv 
\frac{dP_{{\rm M}}^{\mu}}{d\tau}+e{\rm Tr}(\boldsymbol{B}\boldsymbol{P}^{\mu}).
\nonumber 
\end{align}
which satisfy covariance with respect to higher 
gauge transformations as 
\begin{align}
&\delta_{HL}\Bigl(\frac{D'X^{\mu}}{D\tau}\Bigr)
=\Bigl(0, i[\boldsymbol{H}, \frac{D'\boldsymbol{X}^{\mu}}{D\tau}]+
\boldsymbol{L}\frac{dX_{{\rm M}}^{\mu}}{d\tau}\Bigr), \nonumber \\
&\delta_{HL}\Bigl(\frac{D'P^{\mu}}{D\tau}\Bigr)=
\Bigl(-{\rm Tr}\Bigl(\boldsymbol{L}\frac{D'\boldsymbol{P}^{\mu}}{D\tau}\Bigr), i[\boldsymbol{H}, \frac{D'\boldsymbol{P}^{\mu}}{D\tau}]\Bigr). 
\nonumber 
\end{align}
The primes ($'$) being put on these expressions    
indicate that these definitions are not yet final ones, since we have to 
extend our higher transformations further, in order to take into account the negative metric in the covariant 
Poisson brackets. 

\begin{wrapfigure}{r}{6cm}\includegraphics[width=5.8cm,clip]{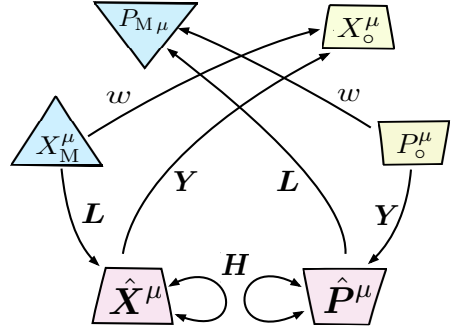}\caption{{\small  Schematic structure of the higher gauge symmetries: The different shapes of the objects indicate different scaling dimensions of canonical variables. The directions of arrows indicate how the variables are  mixed into others (or into themselves) by gauge transformations. The row in the middle represents conserved vectors, while the top rowrepresents the corresponding passive variables. }}\end{wrapfigure}

To understand the necessity of still further extension of gauge 
symmetry, let us reconsider how the covariant mass-shell condition 
\eqref{covmassshell} 
should be generalized to our case. As in the standard treatment 
of covariant relativistic 
particle mechanics, reparametrization 
invariance with respect to $\tau$ will lead automatically, through 
the variation $\delta e$ of the ein-bein auxiliary field, to 
the mass-shell condition for the center-of-mass momentum 
$P_{\circ}^{\mu}\equiv {\rm Tr}\boldsymbol{P}^{\mu}$. 
The time-like component of the latter is then 
constrained to be fixed by spatial components. 
In our case, however, we have 
matrix momenta $\boldsymbol{P}^{\mu}$ with 
their auxiliary accompaniment $P_{{\rm M}}^{\mu}$, the momentum 
M-variable, 
together with conjugate coordinate variables. 
All of the time-like components of these variables must 
be appropriately eliminated in the physical space 
as a consequence of constraints, coming from
higher gauge symmetries. For this purpose, 
the existence of a single
 higher gauge field $\boldsymbol{B}$ other than $\boldsymbol{A}$ turns out to be 
not sufficient. We need yet another matrix gauge 
transformation, which contributes to a 
shift of matrix momentum in the time-like direction. 
A natural candidate for this is 
\begin{align}
\delta_Y\boldsymbol{P}^{\mu}=P_{\circ}^{\mu}\boldsymbol{Y}, \quad 
\delta_YX_{\circ}^{\mu}=-{\rm Tr}(\boldsymbol{Y}\boldsymbol{X}^{\mu})
\label{ytrans}
\end{align}
with an arbitrary traceless (hermitian) matrix parameter $\boldsymbol{Y}$, in analogy with $\boldsymbol{L}$-transformation exhibited 
in \eqref{HLtransX} and \eqref{HLtransP}. 
In fact, the transformation $\delta_H+\delta_Y$ for the 
momentum variables 
essentially coincides with the Nambu transformation 
using our original bracket using trace \eqref{almy}. 
A peculiarity here is that the $\delta_{H}$ part is 
common to both transformations, and hence 
we cannot define these transformation $\delta_H+\delta_L$ and 
$\delta_H+\delta_Y$ as two independent 
transformations, unless we separate the $\delta_H$ part. 
These two sets of gauge symmetries are somewhat analogous 
to the presence of holomorphic and anti-holomorphic parts of 
conformal symmetries in (closed) string theory. 

Here no attentive reader can fail to notice that 
the previous form of the integral invariant \eqref{momentum2} for momentum 
obviously violates the symmetry under \eqref{ytrans}. This is 
easily remedied by a modification with replacement 
$\hat{\boldsymbol{P}}^{\mu}\rightarrow \hat{\boldsymbol{P}}^{\mu}
-(P_{\circ}^2)^{-1}P_{\circ}^{\mu}(\hat{\boldsymbol{P}}\cdot P_{\circ})$. 
More appropriately, we can introduce an additional 
auxiliary (traceless) matrix variable $\boldsymbol{K}$, 
transforming as 
\[
\delta_Y\boldsymbol{K}=\boldsymbol{Y}
\] 
and rewrite an integral invariant as
\[
\int d\tau \, e\, {\rm Tr}(\boldsymbol{P}-P_{\circ}\boldsymbol{K})^2
=\int d\tau \, e\, \Bigl(
\frac{1}{N}P_{\circ}^2+{\rm Tr}(\hat{\boldsymbol{P}}-
P_{\circ}\boldsymbol{K})^2\Bigr)
\]
The variation with respect to $\boldsymbol{K}$ gives
\begin{align}
P_{\circ}\cdot(\hat{\boldsymbol{P}}-P_{\circ}\boldsymbol{K})=0. 
\label{kconstraint}
\end{align}
We may gauge-fix the $Y$-transformation by choosing a condition, say, 
$\boldsymbol{K}=0$, which would lead to 
a constraint 
\[
P_{\circ}\cdot \hat{\boldsymbol{P}}= 0, 
\]
which serves to eliminate explicitly the time-like component of 
the traceless part of matrix momentum.  
The reader might recall that the situation is similar to the Higgs mechanism (or St\"{u}ckelberg formalism) in 
formulating abelian massive vector gauge field covariantly.  

In terms of the infinitesimal canonical 
generator extending \eqref{canogene}, our postulate 
for higher symmetries now amounts to 
\begin{align}
{\cal C}_{w+Y+H+L}=wP_{\circ}\cdot X_{{\rm M}}+
{\rm Tr}\Bigl(-
(P_{\circ}\cdot\boldsymbol{X})\boldsymbol{Y}+
i\boldsymbol{P}_{\mu}[\boldsymbol{H}, \boldsymbol{X}^{\mu}]+(X_{{\rm M}}\cdot \boldsymbol{P})\boldsymbol{L}
\Bigr).
\nonumber 
\end{align}
where the decomposition $w+Y+H+L$ on the l.h.side should be 
obvious from the corresponding 
order of transformation parameters  
on the r.h.side. Here, we have included 
also the 
first term, $w$-transformation with an arbitrary functions 
$w=w(\tau)$, given by 
\begin{align}
\delta_wX_{\circ}^{\mu}=wX_{{\rm M}}, 
\quad \delta_wP_{{\rm M}}^{\mu}=-wP_{\circ}^{\mu}
\nonumber 
\end{align}
which enable one to shift the time-like component of 
$P_{{\rm M}}^{\mu}$ arbitrarily. 

The Lorentz invariance 
of the present canonical formalism for these symmetries is 
ensured by 
\begin{align}
\{{\cal M}^{\mu\nu}, {\cal C}_{w+Y+H+L}\}=0, 
\end{align}
where 
\begin{align}
{\cal M}^{\mu\nu}\equiv 
X_{{\rm M}}^{\mu}P_{{\rm M}}^{\nu}-X_{{\rm M}}^{\nu}P_{{\rm M}}^{\mu}
+{\rm Tr}(\boldsymbol{X}^{\mu}\boldsymbol{P}^{\nu}-
\boldsymbol{X}^{\nu}\boldsymbol{P}^{\mu})
\label{lorentzgene}
\end{align}
are the generators of Lorentz transformations, 
satisfying the correct Lorentz algebra under the 
Poisson-bracket algebras from which we have started our 
canonical formulation.

Taking into account these extensions of higher-gauge symmetries, we can now present the 
final form of covariant derivatives. The new additional 
gauge fields are denoted by $\boldsymbol{Z}$ and $B$ 
corresponding to $\delta_Y$ and $\delta_w$ transformations, 
respectively, the former of which is again traceless by definition. 
\begin{align}
&\frac{DX_{\circ}^{\mu}}{D\tau}= \frac{dX_{\circ}^{\mu}}{d\tau}-eBX_{{\rm M}}^{\mu}+e
{\rm Tr}(\boldsymbol{Z}\hat{\boldsymbol{X}}^{\mu}), \nonumber \\
&\frac{D\hat{\boldsymbol{X}}^{\mu}}{D\tau}=
\frac{d\hat{\boldsymbol{X}}^{\mu}}{d\tau}+ie[\boldsymbol{A}, 
\boldsymbol{X}^{\mu}]-e\boldsymbol{B}X_{{\rm M}}^{\mu}, \nonumber \\
&\frac{DP_{{\rm M}}^{\mu}}{D\tau}=\frac{dP_{{\rm M}}^{\mu}}{d\tau}
+e{\rm Tr}\bigl((\boldsymbol{B}+B)\boldsymbol{P}^{\mu}\bigr),
%=\frac{dP_{{\rm M}}^{\mu}}{d\tau}+e{\rm Tr}(\boldsymbol{B}\boldsymbol{P}^{\mu})+eB_{\circ}P_{\circ}^{\mu}, 
\nonumber \\
&\frac{D\hat{\boldsymbol{P}}^{\mu}}{D\tau}=
\frac{d\hat{\boldsymbol{P}}^{\mu}}{d\tau}+ie
[\boldsymbol{A}, \boldsymbol{P}^{\mu}]
-e\boldsymbol{Z}P_{\circ}^{\mu}. \nonumber 
\end{align}
The transformation laws of the new gauge fields are 
\begin{align}
&\delta_{HL}B={\rm Tr}(\boldsymbol{L}\boldsymbol{Z}), \nonumber \\
&\delta_{HL}\boldsymbol{Z}=i[\boldsymbol{H}, \boldsymbol{Z}], \nonumber \\
&\delta_{w} B =\frac{1}{e}\frac{dw}{d\tau}, \quad 
\delta_{w} \boldsymbol{Z}=0, \nonumber \\
& \delta_Y B=-{\rm Tr}(\boldsymbol{Y}\hat{\boldsymbol{B}}),\nonumber \\
&\delta_Y \boldsymbol{Z}=\frac{1}{e}\frac{d\boldsymbol{Y}}{d\tau}
+i[\boldsymbol{A}, \boldsymbol{Y}]\equiv \frac{1}{e}
\frac{D\boldsymbol{Y}}{D\tau}. \nonumber 
\end{align}
The scaling transformation of newly introduced gauge fields and 
transformation parameters are
\[
B\rightarrow \lambda^2 B, \quad 
\boldsymbol{Z}\rightarrow \lambda^{-2}\boldsymbol{Z}, 
\quad w\rightarrow \lambda^4w, \quad \boldsymbol{Y}
\rightarrow \boldsymbol{Y}.
\]

Now that we have succeeded to construct a 
canonical formalism of higher symmetry, there is 
a basic canonical gauge invariant, namely, the 
generalized Poincar\'{e} integral, involving first derivatives and 
satisfying the scaling symmetry :
\begin{align}
&\int d\tau \Bigl[
P_{{\rm M}\, \mu}\frac{dX_{{\rm M}}^{\mu}}{d\tau}
+{\rm Tr}\Bigl(
\boldsymbol{P}_{\mu}\frac{D\boldsymbol{X}^{\mu}}{D\tau}\Bigr)
\Bigr]
\nonumber \\
&=
\int d\tau \Bigl[
P_{{\rm M}\, \mu}\frac{dX_{{\rm M}}^{\mu}}{d\tau}
+P_{\circ\, \mu}\frac{DX_{\circ}^{\mu}}{D\tau}+{\rm Tr}\Bigl(
\hat{\boldsymbol{P}}_{\mu}\frac{D\hat{\boldsymbol{X}}^{\mu}}{D\tau}\Bigr)
\Bigr]\label{poincare1}\\
&=-\int d\tau \Bigl[
\frac{DP_{{\rm M}\, \mu}}{D\tau}X_{{\rm M}}^{\mu}
+\frac{dP_{\circ\, \mu}}{d\tau}X_{\circ}^{\mu}
+{\rm Tr}\Bigl(
\frac{D\hat{\boldsymbol{P}}_{\mu}}{D\tau}\hat{\boldsymbol{X}}^{\mu}\Bigr)
\Bigr], \nonumber 
%\label{poincare2}
\end{align}
where in the second line we have separated the center-of-mass part, 
and in the third have made partial integration. 
Note that though we are considering local $\tau$-dependent 
canonical transformations as higher gauge transformations, the generalized 
Poincar\'{e} integral is invariant (up to surface terms) 
because of the presence of gauge field. This is in contrast 
to the usual canonical formalism in which a 
time dependent canonical transformation in general 
induces a shift of Hamiltonian 
by the time derivative of corresponding 
infinitesimal generator. In our case, this shift is now 
compensated for by the transformations of gauge fields. 

We require that the $\tau$-derivatives of dynamical 
variables appear only through this invariant, as it should be 
in any standard canonical (first-order) formalism. Hence, the same 
can be said about gauge fields. 
This means that we have already fixed the forms 
of bosonic parts of all Gaussian constraints in our system. 
By taking infinitesimal variations of the gauge fields, we obtain 
four independent constaints, 
\begin{align}
&\delta \boldsymbol{A} :\quad   [\boldsymbol{P}_{\mu}, \boldsymbol{X}^{\mu}]+\cdots \approx 0, 
\label{agauss}\\
&\delta \boldsymbol{B} : \quad \hat{\boldsymbol{P}}_{\mu} X_{{\rm M}}^{\mu}\approx 0, 
\label{Bgauss}\\
&\delta\boldsymbol{Z} : \quad \hat{\boldsymbol{X}}_{\mu}P_{\circ}^{\mu}\approx 0, 
\label{zgauss}\\
&\delta B : \quad P_{\circ\, \mu}X^{\mu}_{{\rm M}}\approx 0, 
\label{bgauss}
\end{align}
where only the first one has a contribution, denoted by ellipsis, from 
fermionic part which we will fix later after 
discussing supersymmetry. All these 
constraints are regarded as ``weak equations" before 
gauge fixing: it is easy 
to check that the algebra of these constraints 
close by themselves, which are therefore of first-class.  
Note that the matrix constraints \eqref{agauss}$\sim$ \eqref{zgauss} 
are all traceless, due to the fact that 
all matrix gauge fields are traceless. It should also be 
noted that if we take into account the equation \eqref{kconstraint} 
as a constraint, it should be treated as 
a second-class constraint, reflecting again that 
it is a sort of gauge-fixing condition for the 
$Y$-gauge transformations, similarly as in 
the case of massive abelian gauge field.  

Since we are supposing a 
flat 11 dimensional Minkowskian spacetimes, we require translation 
invariance under 
$
X_{\circ}^{\mu}\rightarrow X_{\circ}^{\mu}+c^{\mu}
$
with an arbitrary constant vector $c^{\mu}$.  Thus we have
conservation of total momentum
\[
\frac{dP_{\circ}^{\mu}}{d\tau}=0.
\]
As an additional condition, we demand that the system has also 
a translation symmetry with respect to a shift of the 
auxiliary momentum $P_{{\rm M}}^{\mu}$, 
$
P_{{\rm M}}^{\mu}\rightarrow P_{{\rm M}}^{\mu}+b^{\mu}
$
with an arbitrary constant vector $b^{\mu}$, thereby
$X^{\mu}_{{\rm M}}$ being also conserved,
\[
\frac{dX_{{\rm M}}^{\mu}}{d\tau}=0.
\]
Both these symmetries are satisfied by all integral invariants 
discussed so far. 
\begin{wrapfigure}{r}{3.3cm}\vspace{-0.5cm}
\includegraphics[width=3cm,clip]{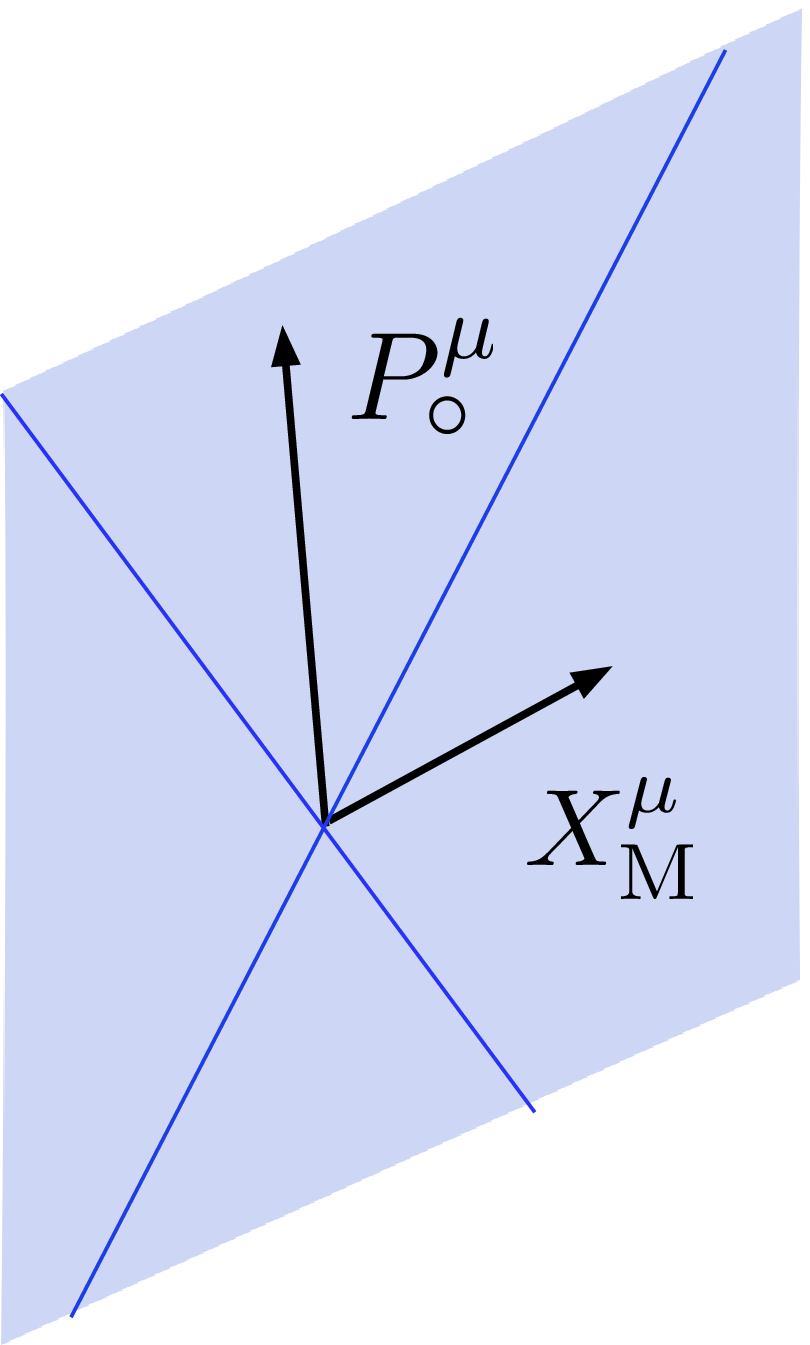}{\small {\caption M-plane spanned by $P_{\circ}^{\mu}$ and $X_{{\rm M}}^{\mu}$}}\end{wrapfigure} 

The conserved center-of-mass momentum $P_{\circ}^{\mu}$ must be time-like (including a possible special case of light-like limit), 
$P_{\circ}^2<0$. Due to the Gauss constraint \eqref{bgauss}, 
this implies that $X_{{\rm M}}^{\mu}$ is a (conserved)  
space-like vector. Thus given an initial condition, we are 
automatically specifying a conserved two-dimensional 
plane spanned by $P_{\circ}^{\mu}$ and 
$X_{{\rm M}}^{\mu}$ in the Minkowski spacetime. 
In the following, we call this plane ``M-plane" for 
convenience. 
In fixing the gauge for higher symmetries, the M-plane will play a preferential 
role, in the sense that there are 
no local physical degrees of freedom living solely on the M-plane. 
The emergence of preferential frame is essentially the same as in any Lorentz covariant 
formulation of particles in configuration space: recall that, given any 
state in a 
many-body system, we have a particular preferential 
frame, namely, the center-of-mass frame, where all of the spatial components of $P_{\circ}^{\mu}$ vanish. Namely, 
the preferential frames appear whenever we 
consider a particular state of particles, which specifies a particular 
configuration of particles.  Only difference in our case 
is that there are two vectors, one time-like and the other space-like, instead of one time-like vector 
in cases of the usual many-body systems. Covariance in the configuration space of particles is guaranteed 
by the existence of generators of Lorentz transformation which 
operate in the space of states and satisfy the correct 
Lorentz algebra. 

We will shortly see that 
the conserved auxiliary vector $X_{{\rm M}}^{\mu}$ 
plays a fundamental role of fixing M-theory scales 
as reviewed in the first part of this lecture. 
It also plays a crucial role in realizing 
supersymmetry in a most 
economical manner in our covariant formulation 
of Matrix theory.

\section{The action of covariantized Matrix theory: bosonic part}

Now we are in a position to write down the 
(bosonic part of the) action of our covariant Matrix theory:
\begin{align}
A_{{\rm boson}}&=\int d\tau 
\Bigl[
P_{\circ}\cdot \frac{DX_{\circ}}{D\tau}
+P_{{\rm M}}\cdot\frac{dX_{{\rm M}}}{d\tau}
%+P_{\circ}\cdot \frac{dX_{\circ}}{d\tau}
+{\rm Tr}\Bigl(
\hat{\boldsymbol{P}}\cdot\frac{D\hat{\boldsymbol{X}}}{D\tau}\Bigr)
%+{\rm Tr}(\boldsymbol{Z}P_{\circ}\cdot\hat{\boldsymbol{X}})
\nonumber \\
&-\frac{e}{2N}P_{\circ}^2
-\frac{e}{2}\, {\rm Tr}(\hat{\boldsymbol{P}}-P_{\circ}\boldsymbol{K})^2 
+ \frac{e}{12} \bigl< [X^{\mu},X^{\nu},X^{\sigma}]
[X_{\mu},X_{\nu},X_{\sigma}]\bigr>
\Bigr].
\label{bosonaction}
\end{align}
The relative normalization between the kinetic momentum 
part and the last potential term is actually arbitrary, since 
it can be freely changed by redefinitions,  
$(X_{{\rm M}}, P_{{\rm M}})\rightarrow (\rho X_{{\rm M}}^{\mu}, 
\rho^{-1}P_{{\rm M}}^{\mu}), (B, \boldsymbol{B})
\rightarrow \rho^{-1}(B, \boldsymbol{B})$, keeping other terms intact. 
This form of the bosonic action
 is characterized by the following four kinds of symmetries. 
\begin{enumerate}
\item Local reparametrization invariance with respect to $\tau$.
\item Global translation invariance with respect to 
$X_{\circ}^{\mu}\rightarrow X_{\circ}^{\mu}+c^{\mu}$ and 
$P_{{\rm M}}^{\mu}\rightarrow P_{{\rm M}}^{\mu}+b^{\mu}$. 
\item Global scaling symmetry \eqref{scaling} under 
$\tau\rightarrow \lambda^2\tau$.  
\item Gauge symmetries under $\delta_{H}+\delta_{L}+\delta_Y+\delta_w$. 
\end{enumerate}
The local symmetries (1) and (4) give constraints. 
The Gauss constraints corresponding to the latter are 
already explicated in the previous section. 
The mass-shell condition corresponding to (1) is
\begin{align}
P_{\circ}^2+{\cal M}_{{\rm boson}}^2\approx 0
\end{align}
with the effective squared-mass
\begin{align}
{\cal M}^2_{{\rm boson}}=N {\rm Tr}(\hat{\boldsymbol{P}}-P_{\circ}\boldsymbol{K})^2
- \frac{N}{6} \bigl< [X^{\mu},X^{\nu},X^{\sigma}]
[X_{\mu},X_{\nu},X_{\sigma}]\bigr>, 
\label{bosonmass}
\end{align}
where the equality is valid only in conjunction with the Gauss-law 
constraints \eqref{agauss}$\sim$\eqref{bgauss}.  This is indicated by the symbol $\approx$: remember that, when a variation of the ein-bein $e(\tau)$ is made, there are contributions from the 
covariant derivatives, involved in the generalized Poincar\'{e} invariant,  which are {\it linear} with respect to 
all the gauge fields and consequently are linear combinations of 
the Gauss constraints. 
It is to be noted that in the large $N$ limit, we are 
interested in the regime where the spectrum of the squared mass 
is of order one and hence is 
independent of $N$ in the large $N$ limit. 

That the effective mass is governed by the internal dynamics of 
this system is ensured by the fact that \eqref{bosonmass} involves only 
traceless matrix variables. 
It is easy to check that the equations of motion preserve the 
Gauss constraints, and hence they are consistently 
implemented. With respect to different roles of 
dynamical variables, it is to be noted that there is no inertial 
kinetic term for the ``M-variables" $(X_{{\rm M}}^{\mu}, P_{{\rm M}}^{\mu})$, due to the symmetry (2). Correspondingly, 
they do not participate to the dynamics actively: 
$X_{{\rm M}}^{\mu}$ is conserved, while 
$P_{{\rm M}}^{\mu}$ is passively determined 
by other variables through 
\[
\frac{DP_{{\rm M}}}{D\tau}=-\frac{\partial}{\partial_{{\rm M}\, \mu}}
{\cal V}
\]
where $-{\cal V}$ is the potential term in the above action and 
hence does not involve $P_{{\rm M}}^{\mu}$. The same can be 
said for the center-of-mass coordinate $X_{\circ}^{\mu}$ 
with respect to its passive character. 

Now our next task is to confirm that this action leads to 
the same results as the light-front Matrix theory 
if we fix the gauge of higher-gauge symmetries 
appropriately and make explicit the condition of 
compactification. We can 
first choose the M-plane spanned by 
$P_{\circ}^{\mu}$ and $X_{{\rm M}}^{\mu}$. 
Since the former can be assumed to be time-like for 
generic states, while the latter then to be space-like due to the 
Gauss constraint \eqref{bgauss}, there is always a 
Lorentz frame where only non-zero components of these 
two conserved vectors are $P_{\circ}^0, P_{\circ}^{10}$ and 
$X_{{\rm M}}^0, X_{{\rm M}}^{10}$, respectively. 
Thus the M-plane is described by 
the light-like components $P_{\circ}^{\pm}\equiv P_{\circ}^{10}
\pm P_{\circ}^0, X_{{\rm M}}^{\pm}\equiv X_{{\rm M}}^{10}
\pm X_{{\rm M}}^0$. We can then impose a gauge condition 
\begin{align}
\hat{\boldsymbol{X}}^+=0, 
\end{align}
using the $\delta_L$-transformation, by which \eqref{zgauss} 
is reduced to
\begin{align}
0=P_{\circ}^+\hat{\boldsymbol{X}}^-+P_{\circ}^-\hat{\boldsymbol{X}}^+=
P_{\circ}\hat{\boldsymbol{X}}^-\quad 
\Rightarrow \hat{\boldsymbol{X}}^-=0
\end{align}
since $P^+_{\circ}\ne 0$ due to our assumption that 
$P_{\circ}^{\mu}$ is time-like. 
With respect to $\delta_Y$-transformations, we choose 
$\boldsymbol{K}=0$ as discussed in the previous section. 
Then the first-order equations of motion 
allow us to express the light-like components 
of the matrix momentum as
\begin{align}
\hat{\boldsymbol{P}}^{\pm}=
\frac{1}{e}\frac{d\hat{\boldsymbol{X}}^{\pm}}{d\tau}+
i[\boldsymbol{A}, \hat{\boldsymbol{X}}^{\pm}]-
\boldsymbol{B}X^{\pm}_{{\rm M}}
\end{align}
which give
\[
\boldsymbol{P}^{\pm}=-\boldsymbol{B}X^{\pm}_{{\rm M}}.
\]
Then, from the Gauss constraint \eqref{Bgauss} 
we obtain
\begin{align}
0=X_{{\rm M}}^-\cdot\hat{\boldsymbol{P}} \quad \Rightarrow \quad \boldsymbol{B}X_{{\rm M}}^2=0 \quad \boldsymbol{B}=0 
\quad \Rightarrow \hat{\boldsymbol{P}}^{\pm}=0. 
\end{align}
Thus all of the light-like traceless matrices vanish in this 
gauge choice. Consequently, the squared mass and 
the remaining Gauss constraint \eqref{agauss} reduce, respectively, 
to
\begin{align}
&{\cal M}^2_{{\rm boson}}=N{\rm Tr}
\Bigl(\hat{\boldsymbol{P}}_i^2 -\frac{1}{2}X_{{\rm M}}^2
[\boldsymbol{X}_i, \boldsymbol{X}_j]^2\Bigr)=\hat{H},\\
&[\boldsymbol{X}_i, \boldsymbol{P}_i]=0, 
\end{align}
which coincides 
with \eqref{lfhamiltonian} of section 3, by 
identifying the Lorentz invariant length of the 
M-variable after recovering the original 
unit of length, as 
\begin{align}
X_{{\rm M}}^2=\frac{1}{\ell_{11}^6}
\end{align}
in terms of the fundamental scale of M-theory. 
This implies that the scaling symmetry is 
broken by this choice. 
We will discuss about the meaning 
of this later. 

In this gauge, the equations of motion for 
the center-of-mass variables and for $X_{{\rm M}}$ are
\[
P_{\circ}^{\pm}=N\Bigl(\frac{dX_{\circ}^{\pm}}{ds}-BX_{{\rm M}}^{\pm}\Bigl), 
\quad 
\frac{dP^{\pm}_{\circ}}{ds}=0, \quad 
\frac{dX_{{\rm M}}^{\pm}}{ds}=0, 
\]
where we defined the re-parametrization invariant 
time parameter $s$ by $ds=ed\tau$. 
By choosing the gauge condition $B=0$ for the 
$\delta_w$-transformation, we have the standard 
form 
\[
P_{\circ}^{\pm}=N\frac{dX_{\circ}^{\pm}}{ds}, 
\]
or 
\begin{align}
X_{\circ}^+=\frac{P_{\circ}^+}{N}s. 
\end{align}
It is compulsory to assume that the relation between the target time 
and the invariant proper time $s$ is 
independent of $N$, as it should be since the systems with different sizes of matrices can always be regarded as subsystems of 
 larger systems with increasingly larger $N$. Otherwise, 
we cannot consistently decompose a given system 
as a composite of subsystems: time $\tau$ or $s$ must 
be common to subsystems which are all 
synchronized with a single global internal time, 
as we have stressed in the previous section as 
a premiss of our canonical approach using a 
single proper time. 
Thus we must have 
\begin{align}
P_{\circ}^+=\frac{2N}{R}
\end{align}
with $R$ being a constant
 parameter which is independent of $N$ but 
can be varied continuously for different choices of the Lorentz frame. 
This somewhat remarkable result is consistent with 
the light-front Matrix theory as an effective theory of 
D0-branes where all D0-branes are supposed to have 
a single quantized unit of KK momentum in the limit of 
small $R$ identified with $R_{11}$. 

Finally, we can derive an effective action for the 
remaining transverse variables by 
substituting 
\[
P_{\circ}^-=-\frac{\hat{H}}{P_{\circ}^+}
\]
back into the original action. The result is, making conversion to the 
second-order formalism after eliminating the momenta,  
\[
A_{{\rm eff}}=\int dx^+\frac{1}{2R}{\rm Tr}\Bigl(\frac{D\hat{\boldsymbol{X}}^i}{Dx^+}\frac{D\hat{\boldsymbol{X}_i} }{Dx^+}+ \frac{R^2}{2\ell_{11}^6}[\boldsymbol{X}_i, \boldsymbol{X}_j][\boldsymbol{X}_i, \boldsymbol{X}_j]\Bigr)
\]
where we redefine the light-like time by $s=2Nx^+/P_{\circ}^+$ $(X_{\circ}=2x^+$). 

The equations of motion for the center-of-mass coordinates and 
momenta does not prohibit us from imposing the BFSS condition 
\[
P_{\circ}^{10}=\frac{N}{R_{11}},
\]
instead of the DLCQ scheme. In this case, we solve the 
mass-shell constraint as 
\[
P_{\circ}^{0}%={\cal M_{{\rm trans}}}
=
\sqrt{(P_{\circ}^{10})^2+N{\rm Tr}\Bigl(\hat{\boldsymbol{P}}_i\cdot\hat{\boldsymbol{P}} _i
- \frac{1}{2}
X_{{\rm M}}^2
[\boldsymbol{X}_i,\boldsymbol{X}_j][\boldsymbol{X}_i,\boldsymbol{X}_j]
\Bigr)}.
\]
Then the effective action is 
\[
A_{{\rm spat\, boson}}=
\int dt \Bigl[{\rm Tr}\Bigl(
\hat{\boldsymbol{P}}_i\frac{D\hat{\boldsymbol{X}}_i}{Dt}\Bigr)
-P_{\circ}^0
\Bigr]
\]
with the time parameter $t=X_{\circ}^0=P_{\circ}^{10}s/N=s/R_{11}$. 
By eliminating $\hat{\boldsymbol{P}}_i$ in terms 
of the coordinate variables, we obtain the following Born-Infeld-like 
action:
\[
A_{{\rm spat\, boson}}=-
\int dt\,  {\cal M}_{{\rm spat}}\sqrt{N}\Bigl[
1-\frac{1}{N}{\rm Tr}\Bigl(\frac{D\hat{\boldsymbol{X}}_i}{Dt}
\frac{D\hat{\boldsymbol{X}}_i}{Dt}\Bigr)\Bigr]^{1/2},
\]
\[
{\cal M}_{{\rm spat}}\equiv \Bigl[\frac{N}{R_{11}^2}
-\frac{1}{2\ell_{11}^6}{\rm Tr}\bigl([\boldsymbol{X}_i,\boldsymbol{X}_j]
[\boldsymbol{X}_i,\boldsymbol{X}_j]\bigr)\Bigr]^{1/2}. 
\]
If we assume that the kinetic term, 
${\rm Tr}\Bigl(\frac{D\hat{\boldsymbol{X}}_i}{Dt}
\frac{D\hat{\boldsymbol{X}}_i}{Dt}\Bigr)$,  and 
the potential term, ${\rm Tr}\bigl([\boldsymbol{X}_i,\boldsymbol{X}_j]
[\boldsymbol{X}_i,\boldsymbol{X}_j]\bigr)$, 
are at most of order $O(1)$ with respect to $N$, the above effective 
action is approximated as
\[
\int dt\, \frac{N}{R_{11}}\Bigl[-1
+
\frac{1}{2N}{\rm Tr}\Bigl(\frac{D\hat{\boldsymbol{X}}_i}{Dt}\frac{D\hat{\boldsymbol{X}}_i}{Dt}
+
\frac{R_{11}^2}{2\ell_{11}^6}[\boldsymbol{X}_i,\boldsymbol{X}_j]
[\boldsymbol{X}_i,\boldsymbol{X}_j] \Bigr)+O(\frac{1}{N^2})\Bigr].
\]
Of course, this is consistent with a natural expectation 
from our viewpoint on the relationship between 
the IMF and DLCQ schemes, discussed in section 3. 
On the other hand, our result shows that in the opposite limit 
$R_{11}\rightarrow \infty$ with fixed $N$, the system 
becomes a very peculiar and singular system which does not 
have standard kinetic terms. 

At this juncture, let us consider the meaning of the violation 
of scaling symmetry, which is required in order to relate our system 
with light-front Matrix theory. Namely, 
the 11 dimensional Planck length emerges by specifying 
the value of $X_{{\rm M}}^2$ as an initial condition. 
This determines the coupling constant for the 
internal dynamics of the system. A natural interpretation of 
this situation seems that $X_{{\rm M}}^2$ defines a 
super-selection rule with respect to scaling transformations. 
Namely, once its value is fixed by initial condition, no superposition 
is allowed among different values of $X_{{\rm M}}^2$. 
The scale symmetry means that any two systems 
with different values of $X_{{\rm M}}^2$ are 
 mapped  into each other with 
a simple rescaling of dynamical variables. 
Thus all the different super-selection sectors actually 
describe essentially the same physics, apart from 
global scaling transformations. The initial condition 
 just selects one of the continuously distributed super-selection sectors. In this sense, scale symmetry is 
spontaneously broken. On the other hand, 
scale symmetry signifies an important fact that 
our theory has one and only one fundamental length scale $\ell_{11}$ 
through spontaneous symmetry breaking. 

It should be noted also that, even though states are not 
superposed between different values of invariant $X_{{\rm M}}^2$,  states with different Lorentz components of $X_{{\rm M}}^{\mu}$ must be 
allowed to be superposed. That this is the case is seen, for instance,  from 
the constraint \eqref{bgauss}, which in the light-like 
coordinates of the M-plane takes the form, 
\[
P_{\circ}^-%=-\frac{P_{\circ}^+X_{\uhdb}^-}{X_{\uhdb}^+}=
%-\frac{P_{\circ}^+}{(X_{\uhdb}^+)^2}(X_{\uhdb}^2-(X^i_{\uhdb})^2)
=-\frac{P_{\circ}^+}{(X_{{\rm M}}^+)^2}X_{{\rm M}}^2
\qquad 
\mbox{or} 
\qquad
P_{\circ}^0%=\frac{P_{\circ}^{10}X_{\uhdb}^{10}}{X_{\uhdb}^0}
=\frac{P_{\circ}^{10}X_{{\rm M}}^{10}}{\sqrt{(X_{{\rm M}}^{10})^2-X_{{\rm M}}^2}}.
\]
This implies that different states with different ``energies"$P_{\circ}^-$ 
or $P_{\circ}^{10}$, in general, have different values 
$X_{{\rm M}}^+$ or $X_{{\rm M}}^{10}$, respectively. 
The states with different energies are certainly 
superposed in quantum dynamics, and hence also states with 
different values of these components of the M-variable with 
fixed $X_{{\rm M}}^2$ are in general superposed. 
Incidentally, these relations show that the light-like 
limit $P_{\circ}^-
\rightarrow 0$ corresponds to a singular limit 
$X_{{\rm M}}^+\rightarrow \infty$ or $X_{{\rm M}}^{10} 
\rightarrow \infty$. 

\section{Supersymmetry}
Now let us come to our last subject of this lecture. The question 
we have to ask is now whether and how our covariantized 
Matrix theory can have supersymmetry, which is also one of 
indispensable elements ensuring the compatibility 
of this system with eleven-dimensional gravity. 
One of obstacles in formulating supersymmetry 
in a covariant fashion is that we have to 
reduce the number of degrees of freedom 
associated with the fermion variables in 
11 dimensions: a single Majonara fermion has 
32 (real) components. If we suppose that supersymmetry 
is realized without spontaneous symmetry breaking, 
the number of physical degrees of freedom must match 
between bosonic and fermionic degrees of freedom. 
The bosonic coordinate degrees of freedom is 
$11-2-1=8$ for each real components 
of matrices in our system: $-2$ corresponds to 
higher gauge symmetry and $-1$ to 
the ordinary SU($N$) gauge symmetry. Thus, 32 of the 
Majorana fermion must be reduced to $2\times 8=16$. 
In the classical theory of supermembrane, 
this reduction is made due to the presence of 
a fermionic gauge symmetry, the so-called $\kappa$-symmetry. 
It is also the case of manifestly covariant formulations of 
superstrings in 10 dimensions. 
Using this $\kappa$-symmetry, we can put a gauge condition 
on fermions, achieving the required reduction. 
In the same vein, we may try to 
find some extension of $\kappa$-like fermionic gauge symmetry 
in our system. However, for our purpose it is 
sufficient if we have some way of imposing 
condition of reduction directly without violating Lorentz covariance. 
In that case, the existence of such fermionic 
gauge symmetry is not a necessary prerequisite for a 
covariant formulation of supersymmetry. In this lecture, 
we take this standpoint. Of course, this does {\it not} mean that the 
fermionic gauge symmetry is impossible: at least from 
esthetic viewpoint, such a symmetry would be still desirable, 
though from a practical viewpoint it may neither be necessary nor  
useful. One of the reasons for our standpoint is that in the case of fermionic variables, it is impossible to separate them into 
the coordinate and momentum variables in a 
covariant fashion, since 
they are inextricably mixed under Lorentz transformations, 
corresponding to their first-order 
nature of the dynamics of fermions. This feature necessarily leads to second-class constraints 
in the canonical formalism, and compels us 
to assign the same transformation law for all 
spinor components on an equal footing, which can be 
satisfied only for the usual SU($N$) gauge transformation. 

We denote the fermionic Majonara variable by $\Theta_{\circ}$ and 
$\boldsymbol{\Theta}$ : the former  is the 
fermionic partner to the center-of-mass bosonic variables 
$(X_{\circ}^{\mu}, P_{\circ}^{\mu})$. The latter is a {\it traceless}  hermitian 
matrix whose real components are Majorana fermions separately. 
For notational brevity, we suppress the $\hat{\,\, }$ symbol 
for the fermion traceless matrix. 
One basic assumption corresponding to our standpoint 
explained above is that there is no fermionic counterpart for 
the bosonic M-variables $(X_{{\rm M}}^{\mu},P_{{\rm M}}^{\mu})$.  
This implies that the fermionic 
variables do not subject to higher-gauge transformations:
\begin{align}
&\delta_{HL}\Theta=\delta_{HY}\Theta =\delta_H\Theta=
(0, i\sum_r[F^{(r)},G^{(r)}, \Theta])=(0,\delta_H \boldsymbol{\Theta}), 
\nonumber \\
&\delta_H\boldsymbol{\Theta}=i[\boldsymbol{H},\boldsymbol{\Theta}]
\end{align}
Consequently, as for the invariants involving only 
fermionic variables, we can adopt usual trace of 
product of matrices. For expressing invariants involving 
both fermionic and bosonic variables, 3-bracket 
notation is still necessary and useful. 

We first treat the center-of-mass part, which 
can be regarded as if it is a single {\it massive }relativistic particle. 
It is then natural to define its action 
just by adopting the standard formulation of 
a single relativistic superparticle,\cite{siegel} as 
\begin{align}
\int d\tau P_{\circ\, \mu}\bar{\Theta}_{\circ}\Gamma^{\mu}
\frac{d\Theta_{\circ}}{d\tau}, 
\end{align}
which is obtained from the bosonic Poincar\'{e} invariant 
by a replacement,
\[
\frac{dX^{\mu}_{\circ}}{d\tau}
\rightarrow \frac{dX^{\mu}_{\circ}}{d\tau}+
\bar{\Theta}_{\circ}\Gamma^{\mu}
\frac{d\Theta_{\circ}}{d\tau}.
\]
Corresponding to this origin of the fermionic action, 
the center-of-mass system is supersymmetric under 
\begin{align}
\delta_{\varepsilon}\Theta_{\circ}=-\varepsilon, 
\quad \delta_{\varepsilon}X_{\circ}^{\mu}=\bar{\varepsilon}
\Gamma^{\mu}\Theta_{\circ}, 
\quad 
\delta_{\varepsilon}P_{\circ}^{\mu}=0,
\end{align}
satisfying 
\[
\delta_{\varepsilon}\Bigl(  \frac{dX^{\mu}_{\circ}}{d\tau}+
\bar{\Theta}_{\circ}\Gamma^{\mu}
\frac{d\Theta_{\circ}}{d\tau}\Bigr)=0,
\]
with a constant $\varepsilon$. $\Gamma^{\mu}$'s are 
11 dimensional Dirac matrices in the Majorana representation. 
This transformation is not a linear transformation: it is 
characterized by the shift-type 
transformation of $\Theta_{\circ}$, signifying 
that $\Theta_{\circ}$ is super-coordinate 
accompanying the bosonic coordinates $X_{\circ}^{\mu}$. 
The bosonic M-variables and 
all traceless matrix variables are inert under this 
supersymmetry. 
Note that the scaling transformation for the fermionic 
variables are $\Theta_{\circ}\rightarrow \lambda^{1/2}\Theta_{\circ}, 
\varepsilon\rightarrow \lambda^{1/2}\varepsilon$, and also that the $\delta_w$ 
symmetry of the bosonic part is not spoiled: it is still valid 
with the covariant derivative $DX_{\circ}^{\mu}/D\tau$ and the 
conservation law of $X_{{\rm M}}^{\mu}$. Thus the 
Gauss law \eqref{bgauss} is intact, being invariant under 
the above supersymmetry transformation. 

The equation of motion for $\Theta_{\circ}$ is 
\[
P_{\circ}\cdot \Gamma\frac{d\Theta_{\circ}}{d\tau}=0
\]
which for generic time-like $P_{\circ}^{\mu}$ leads to the 
conservation law
\[
\frac{d\Theta_{\circ}}{d\tau}=0. 
\]
Thus the on-shell equations of motion for bosonic 
center-of-mass coordinates are not modified. 

The generic quantum states consist of 
fundamental massive super-multiplet of dimensions $2^{16}$. 
In the special limit of light-like center-of-mass satisfying 
$P_{\circ}^2=0$, it is well known that this system has 
a local fermionic symmetry called Siegel symmetry 
which is the origin of more general $\kappa$-symmetry 
of string and membrane theories. 
\[
\delta_{\kappa}\Theta_{\circ}=P_{\circ}\cdot \Gamma \kappa, \quad 
\delta_{\kappa}X_{\circ}^{\mu}=-\bar{\Theta}_{\circ}\Gamma^{\mu}
\delta_{\kappa}\Theta_{\circ}, 
\]
where $\kappa=\kappa(\tau)$ is an arbitrary Majorana 
spinor function. This allows one to eliminate 
a half of components of $\Theta_{\circ}$ adjoined with 
a suitable redefinition of $X_{\circ}^{\mu}$. Thus, 
a massless 
graviton multiplet consists of $2^{16/2}=2^8$ states. 
But in the present case, we are in general dealing with many-body 
states of such gravitons which obeys the massive representations, 
where $-P_{\circ}^2>0$. 

Now we turn to traceless matrix part, describing the 
internal dynamics of the system. Unlike the center-of-mass 
case, supersymmetry transformations of traceless matrices 
are expected to start from a linear form without 
shift-type contributions, but with possible 
nonlinear corrections of higher orders. 
As for the bosonic coordinates, we start from 
\begin{align}
\delta_{\epsilon}\hat{\boldsymbol{X}}^{\mu}=
\bar{\epsilon}\Gamma^{\mu}\boldsymbol{\Theta}, 
\label{dynasusyX}
\end{align}
where we have different symbol $\epsilon$ for the fermionic 
parameter of transformation, in order to keep in mind that 
this transformation is independent of the 
previous one for the center-of-mass system. 
We call this supersymmetry dynamical 
supersymmetry. For this type of 
transformations to be successfully formulated, 
as we have discussed in the beginning of this section, 
we have to impose some constraints, thereby which 
the degrees of freedom match between bosonic and 
fermionic sides. This necessarily comes about by 
requiring that supersymmetry transformation should 
keep the bosonic Gauss constraints consistently. 
If we assume naturally that $P_{\circ}^{\mu}$ and 
$X_{{\rm M}}^{\mu}$ are inert under dynamical 
super transformation, the constraints, \eqref{Bgauss} and 
\eqref{zgauss}, require
\begin{align}
&X_{{\rm M}}\cdot \delta_{\epsilon}\hat{\boldsymbol{P}}=0, \quad P_{\circ}\cdot \delta_{\epsilon}\hat{\boldsymbol{X}}=0, 
\label{susyreq}
\end{align}
respectively.  There is a natural 
projection condition suitable for our demand, 
due to the existence of the M-plane in the bosonic sector. 
We define (real) projection operators 
\begin{align}
P_{\pm}\equiv \frac{1}{2}(1\pm \Gamma_{\circ}\Gamma_{{\rm M}}), 
\quad 
P_{\pm}^2=P_{\pm}, \quad P_{\pm}P_{\mp}=0
\end{align}
where 
\[
\Gamma_{{\rm M}}\equiv \frac{X_{{\rm M}}\cdot \Gamma}{\sqrt{X_{{\rm M}}^2}}, 
\quad 
\Gamma_{\circ}\equiv \frac{P_{\circ}\cdot \Gamma}{
\sqrt{-P_{\circ}^2}}
\]
are conserved and Lorentz invariant, satisfying 
\begin{align}
&\Gamma_{{\rm M}}\Gamma_{\circ}+\Gamma_{\circ}\Gamma_{{\rm M}}=0, 
\quad 
\Gamma_{{\rm M}}^2=1, \quad \Gamma_{\circ}^2=-1, 
\quad 
(\Gamma_{\circ}\Gamma_{{\rm M}})^2=1, \nonumber \\
& P_+\Gamma_{{\rm M}}=\Gamma_{{\rm M}}P_-, \,\, 
P_+\Gamma_{\circ}=\Gamma_{\circ}P_-, \quad 
P_{\pm}\Gamma_i=\Gamma_iP_{\pm}, 
\label{projrelation}
\end{align}
where $i$ denotes SO(9) directions in 
any (orthogonal) basis, being transverse to 
the M-plane. For the validity of these relations, 
it is crucial to use the orthogonality of two 
conserved vectors $P_{\circ}^{\mu}$ and 
$X_{{\rm M}}^{\mu}$, namely the 
Gauss constraint \eqref{bgauss} associated with 
$\delta_w$-gauge transformations. 
Thus it should be kept in mind that the dynamical 
supersymmetry is satisfied in each sector with 
definite values  of these conserved and 
mutually orthogonal vectors. 

The last relation \eqref{projrelation} shows that 
we can clearly separate the directions between those (called 
``longitudinal") 
along the M-plane and those (called 
``transversal") orthogonal to the M-plane. 
This is precisely what we need in order to meet our requirements 
\eqref{susyreq}. 
We introduce the projection conditions as
\begin{align}
P_-\boldsymbol{\Theta}=\boldsymbol{\Theta}, \, \, P_+\boldsymbol{\Theta}=0, \quad 
\mbox{(or \,\, equivalently \,\, $\bar{\boldsymbol{\Theta}}P_+=
\bar{\Theta},\, \, \bar{\boldsymbol{\Theta}}P_-=0$)} 
\label{projection}
\end{align}
together with the opposite projection on $\epsilon$, 
\begin{align}
P_+\epsilon=\epsilon, \, \, P_-\epsilon=0, 
\quad 
\mbox{(or \,\, equivalently \,\, $\bar{\epsilon}P_-=
\bar{\epsilon},\, \, \bar{\epsilon}P_+=0$)} .
\label{projection2}
\end{align}
This eliminates a half of 32 Majorana components, as 
required. 
Using the postulate \eqref{dynasusyX}, we can confirm that 
the second of \eqref{susyreq} is indeed satisfied:
\begin{align}
\bar{\epsilon}P_{\circ}\cdot \Gamma\boldsymbol{\Theta}
=\bar{\epsilon}P_-(P_{\circ}\cdot \Gamma) P_-
\boldsymbol{\Theta}
=\bar{\epsilon}(P_{\circ}\cdot \Gamma) P_+P_-\boldsymbol{\Theta}=0,
\end{align}
and also 
\begin{align}
\bar{\epsilon}X_{{\rm M}}\cdot \Gamma\boldsymbol{\Theta}
=\bar{\epsilon}P_-(X_{{\rm M}}\cdot \Gamma)P_-\boldsymbol{\Theta}=\bar{\epsilon}X_{{\rm M}}\cdot \Gamma P_+P_-\boldsymbol{\Theta}=0,
\label{Mspinorcond}
\end{align}
while 
\begin{align}
\bar{\epsilon}\Gamma_i\boldsymbol{\Theta}=
\bar{\epsilon}P_-\Gamma_iP_-\boldsymbol{\Theta}
=\bar{\epsilon}\Gamma_iP_-\boldsymbol{\Theta}
=\bar{\epsilon}P_-\Gamma_i\boldsymbol{\Theta}
\end{align}
can be non-vanishing 
for all $i$'s, transverse to both $P_{\circ}^{\mu}$ and $X_{{\rm M}}^{\mu}$. 
Thus as expected, the dynamical supersymmetry 
is effective only for the spacetime directions 
which are transverse to the M-plane. 
This is natural, since as we have seen clearly 
in the previous section that internal 
dynamics is associated entirely to the 
transverse variables. 

In fact, if we adopt the light-like Lorentz frame which 
we have introduced in discussing gauge fixing in 
the previous section, the projection condition is 
equivalent to the ordinary light-cone condition 
for fermionic matrices: we can rewrite \eqref{projection} 
by multiplying $\Gamma_{\circ}$ on both sides, as
\begin{align}
(\Gamma_{\circ}-\Gamma_{{\rm M}})\boldsymbol{\Theta}=0,
\label{covlightcone}
\nonumber 
\end{align}
which reduces to 
\begin{align}
0
&=\frac{1}{2\sqrt{-P_{\circ}^2}}\Bigl(
P_{\circ}^+\Gamma^-+
P_{\circ}^-\Gamma^+
-
\frac{\sqrt{-P_{\circ}^+P_{\circ}^-}}{\sqrt{X_{{\rm M}}^2}}
(X_{{\rm M}}^+\Gamma^-+X_{{\rm M}}^-\Gamma^+)
\Bigr)\boldsymbol{\Theta}\nonumber \\
&=\frac{1}{2\sqrt{-P_{\circ}^2}}\Bigl(
P_{\circ}^+\Gamma^-+
P_{\circ}^-\Gamma^+
-
\frac{P_{\circ}^+}{X_{{\rm M}}^+}
(X_{{\rm M}}^+\Gamma^-+\frac{X_{{\rm M}}^2}{X_{{\rm M}}^+}\Gamma^+)
\Bigr)\boldsymbol{\Theta}\nonumber \\
&=-\sqrt{-
\frac{P_{\circ}^-}{P_{\circ}^+}}\Gamma^+\boldsymbol{\Theta}. 
\nonumber 
\end{align}
We will give full transformation laws for dynamical 
supersymmetry, after showing the total supersymmetric action 
involving both bosonic and fermionic variables in the next section. 

\section{The total supersymmetric action}
The total action is 
\begin{align}
A=A_{{\rm boson}}+A_{{\rm fermion}},
\end{align}
where $A_{{\rm boson}}$ is given by \eqref{bosonaction} and 
\begin{align}
A_{{\rm fermion}}=&
\int d\tau \, \Bigl[\bar{\Theta}_{\circ}P_{\circ}\cdot \Gamma \frac{d\Theta_{\circ}}{d\tau}
+\frac{1}{2}{\rm Tr}\Bigl(
\bar{\boldsymbol{\Theta}}\Gamma_{\circ}
\frac{D\boldsymbol{\Theta}}{D\tau}\Bigr)\nonumber \\
&
-e\frac{i}{4}\bigl<
\bar{\Theta}, \Gamma_{\mu\nu}[X^{\mu}, X^{\nu}, \Theta]\bigr>\Bigr], 
\label{fermiaction}
\end{align}
\begin{align}
&\bigl<
\bar{\Theta}, \Gamma_{\mu\nu}[X^{\mu}, X^{\nu}, \Theta]\bigr>
%&=2{\rm Tr}\bigl(\bar{\boldsymbol{\Theta}}\Gamma_{\mu\nu}X_{\uhdb}^{\mu}[\boldsymbol{X}^{\nu},\boldsymbol{\Theta}]\bigr)\\
\nonumber \\
&=2\sqrt{X_{{\rm M}}^2}{\rm Tr}\bigl(
\bar{\boldsymbol{\Theta}}\Gamma_{\circ}\Gamma_i
[\boldsymbol{X}_i,\boldsymbol{\Theta}]
\bigr)
=2\sqrt{X_{{\rm M}}^2}{\rm Tr}\bigl(
\bar{\boldsymbol{\Theta}}\Gamma_{\circ}\Gamma_{\mu}
[\boldsymbol{X}^{\mu},\boldsymbol{\Theta}]
\bigr). \label{fermipotential}
\end{align}
The last expression of the femionic potential terms is 
derived by using
\[
\bigl<
\bar{\Theta}, \Gamma_{\mu\nu}[X^{\mu}, X^{\nu}, \Theta]\bigr>=2{\rm Tr}\bigl(\bar{\boldsymbol{\Theta}}\Gamma_{\mu\nu}X_{{\rm M}}^{\mu}[\boldsymbol{X}^{\nu},\boldsymbol{\Theta}]\bigr)
\] 
which is rewritten as above due to the 
projection condition $\Gamma_{{\rm M}}\boldsymbol{\Theta}=
\Gamma_{\circ}\boldsymbol{\Theta}$. 
Here it is to be noted that the normalizations of the 
center-of-mass part and of the traceless matrix 
part is different, such that, in the latter, 
the scaling dimensions of $\boldsymbol{\Theta}$ is chosen to be 
zero, while that of the susy parameter $\epsilon$ is 1, 
in order to simplifying the expressions. 
The full dynamical supersymmetry transformations are 
\begin{align}
&\delta_{\epsilon}\hat{\boldsymbol{X}}^{\mu}=\bar\epsilon\Gamma^{\mu}
\boldsymbol{\Theta},\\
&\delta_{\epsilon}\hat{\boldsymbol{P}}_{\mu}=
i\sqrt{X_{{\rm M}}^2}\, \bigl[\bar{\boldsymbol{\Theta}}\Gamma_{\mu\nu}\epsilon, \tilde{\boldsymbol{X}}^{\nu}], \quad 
\delta_{\epsilon}\boldsymbol{K}=0, 
\label{momsusy}
%\boldsymbol{X}^{\nu}-(X_{\uhdb}^2)^{-1}\,X_{\uhdb}^{\nu}(X_{\uhdb}\cdot \boldsymbol{X})\bigr], 
\\
&\delta_{\epsilon}\boldsymbol{\Theta}=P_-\bigl(\Gamma_{\circ}
\Gamma_{\mu}\hat{\boldsymbol{P}}^{\mu}\epsilon
-\frac{i}{2}\sqrt{X_{{\rm M}}^2}\, \Gamma_{\circ}\Gamma_{\mu\nu}
\epsilon[\tilde{\boldsymbol{X}}^{\mu},\tilde{\boldsymbol{X}}^{\nu}]\bigr), \\ 
&\delta_{\epsilon}\boldsymbol{A}=\sqrt{X_{{\rm M}}^2}\, \bar{\boldsymbol{\Theta}}\epsilon, \\
&
\delta_{\epsilon}\boldsymbol{B}=i\bigl(X_{{\rm M}}^2\bigr)^{-1}[\delta_{\epsilon}\boldsymbol{A}, 
X_{{\rm M}}\cdot \boldsymbol{X}], 
\\
&\delta_{\epsilon}\boldsymbol{Z}=i(P_{\circ}^2)^{-1}[\delta_{\epsilon}\boldsymbol{A}, P_{\circ}\cdot \boldsymbol{P}]\nonumber \\
&\hspace{0.7cm}+\frac{X_{{\rm M}}^2}{2P_{\circ}^2}
([\delta_{\epsilon}\boldsymbol{X}^{\mu}, [P_{\circ}\cdot \boldsymbol{X}, 
\boldsymbol{X}_{\mu}]]+
[\boldsymbol{X}^{\mu}, [P_{\circ}\cdot \boldsymbol{X}, 
\delta_{\epsilon}\boldsymbol{X}_{\mu}]]),
%\\
%&\delta_{\epsilon}X_{\circ}^{\mu}=\frac{1}{2\sqrt{-P_{\circ}^2}}\bar{\boldsymbol{\Theta}}\check{\Gamma}^{\mu}\delta_{\epsilon} \boldsymbol{\Theta},
\end{align}
%where we have defined
with
\begin{align}
%&\check{\Gamma}^{\mu}=\Gamma^{\mu}-\frac{P_{\circ}^{\mu}}{P_{\circ}^2}(P_{\circ}\cdot \Gamma)=\Gamma^{\mu}+\frac{P_{\circ}^{\mu}}{\sqrt{-P_{\circ}^2}}\Gamma_{\circ}, \\
&\tilde{\boldsymbol{X}}^{\mu}=
\boldsymbol{X}^{\mu}-
\frac{1}{X_{{\rm M}}^2}X_{{\rm M}}^{\mu}
(\boldsymbol{X}\cdot X_{{\rm M}} )
-\frac{1}{P_{\circ}^2}P_{\circ}^{\mu}(\boldsymbol{X}\cdot 
P_{\circ}).
\end{align}
The existence of gauge fields is crucial for dynamical 
supersymmetry. 
It is easy to check that the transformation law \eqref{momsusy} for 
the momentum matrix satisfies the first of our requirements 
\eqref{susyreq}. 

There is a caveat here : in deriving these transformation laws, we have 
to assume the conservation laws for 
$P_{\circ}^{\mu}$ and $X_{{\rm M}}^{\mu}$ which 
are actually resulting only after using the equations of motion 
for these variables, together with the Gauss law \eqref{bgauss}, 
as we have already alluded to in the previous section. 
I would like to refer the reader to my original paper\cite{yone2} 
for a derivation of these results. 

With this caveat, we can also express the supersymmetry  transformation laws in a form of the algebra of supersymmetry 
generators, using Dirac bracket 
which takes into account the primary second-class 
constraint for the fermion matrices.
 Denoting the 
canonical conjugate to $\boldsymbol{\Theta}$ by $\boldsymbol{\Pi}$, the primary second-class constraint for the traceless fermion matrices is
\begin{align}
\boldsymbol{\Pi}+\frac{1}{2}\bar{\boldsymbol{\Theta}}\Gamma_{\circ}=0, 
\quad (\boldsymbol{\Pi}P_-=\boldsymbol{\Pi})
\end{align}
which satisfy
 the Poisson bracket algebra expressed in a component form
% \footnote{Note that $\{\Pi_{\alpha}^A, \Theta_{\beta}^B\}_{{\rm P}}=(P_-)_{\beta\alpha}\delta^{AB}$. Then, $\{\Pi_{\alpha}^A, (\bar{\Theta}^B\Gamma_{\circ})_{\beta}\}_{{\rm P}}=\delta^{AB}(P_-)_{\gamma\alpha}(\Gamma^0\Gamma_{\circ})_{\gamma\beta}=\delta^{AB}(\Gamma^0\Gamma_{\circ}P_-)_{\beta\alpha}=\delta^{AB}(P_-^{{\rm T}}\Gamma^0\Gamma_{\circ})_{\beta\alpha}$, due to $(\Gamma^0\Gamma_{\circ})_{\beta\alpha}=(\Gamma^0\Gamma_{\circ})_{\alpha\beta}$. }
 \begin{align}\{\Pi_{\alpha}^A+\frac{1}{2}(\bar{\Theta}^A\Gamma_{\circ})_{\alpha}, \Pi_{\beta}^B+\frac{1}{2}
(\bar{\Theta}^A\Gamma_{\circ})_{\beta}\}_{{\rm P}}
=(\Gamma^0\Gamma_{\circ}P_-)_{\alpha\beta}\delta^{AB}, 
\end{align}
where we have denoted the spinor indices by $\alpha, \beta, \ldots, $.  
The indices $A, B, \ldots$ refer to the components 
with respect to the
 traceless spinor matrices using an hermitian orthogonal basis 
$\boldsymbol{\Theta}=\sum_A\Theta^A\boldsymbol{T}^A$ 
satisfying ${\rm Tr}(\boldsymbol{T}^B\boldsymbol{T}^B)=\delta^{AB}
$ of SU($N$) algebra. 
The non-trivial Dirac brackets for traceless 
matrices are then 
\begin{align}
&
%\{\Theta^A_{\alpha}, \Theta^B_{\beta}\}_{{\rm D}}=(P_-\Gamma_{\circ}\Gamma^0)_{\alpha\beta}\delta^{AB}, \quad \mbox{or} \quad 
\{\Theta^A_{\alpha}, 
\bar{\Theta}^B_{\beta}\}_{{\rm D}}=-(P_-\Gamma_{\circ})_{\alpha\beta}
\delta^{AB}, 
\quad 
\\
&\{\hat{X}^A_{\mu}, \hat{P}^B_{\nu}\}_{{\rm D}}=\eta_{\mu\nu}\delta^{AB}.
\end{align}
Then the supercharge 
defined by 
\begin{align}
{\cal Q}=P_-{\rm Tr}(\tilde{\hat{\boldsymbol{P}}}_{\mu}\Gamma^{\mu}\boldsymbol{\Theta}
-\frac{i}{2}\sqrt{X_{{\rm M}}^2}[\tilde{\boldsymbol{X}}^{\mu}, \tilde{\boldsymbol{X}}^{\nu}]\Gamma_{\mu\nu}\boldsymbol{\Theta})
\end{align}
with 
\begin{align}
\tilde{\hat{\boldsymbol{P}}}^{\mu}=
\hat{\boldsymbol{P}}^{\mu}-
\frac{1}{X_{{\rm M}}^2}X_{{\rm M}}^{\mu}
(\hat{\boldsymbol{P}}\cdot X_{{\rm M}} )
-\frac{1}{P_{\circ}^2}P_{\circ}^{\mu}(\hat{\boldsymbol{P}}\cdot 
P_{\circ}),
\end{align}
satisfies
\begin{align}
\{\bar{\epsilon}{\cal Q}, \hat{\boldsymbol{X}}^{\mu}\}_{{\rm D}}&=
-\bar{\epsilon}\Gamma^{\mu}\boldsymbol{\Theta}, \\
\{\bar{\epsilon}{\cal Q}, \hat{\boldsymbol{P}}_{\mu}\}_{{\rm D}}
&=-i\sqrt{X_{{\rm M}}^2}[%\boldsymbol{X}^{\nu}-(X_{{\rm M}}^2)^{-1}X^{\nu}_{{\rm M}}(X_{{\rm M}}\cdot \boldsymbol{X}), 
\bar{\epsilon}
\Gamma_{\mu\nu}\boldsymbol{\Theta}, \tilde{\boldsymbol{X}}^{\nu}]=i
\sqrt{X_{{\rm M}}^2}[\bar{\boldsymbol{\Theta}}\Gamma_{\mu\nu}\epsilon, 
%\boldsymbol{X}^{\nu}-(X_{{\rm M}}^2)^{-1}X^{\nu}_{{\rm M}}(X_{{\rm M}}\cdot \boldsymbol{X})
\tilde{\boldsymbol{X}}^{\nu}],\\
\{\bar{\epsilon}{\cal Q}, \boldsymbol{\Theta}\}_{{\rm D}}&=
-P_-\bigl(\Gamma_{\circ}
\Gamma_{\mu}\hat{\boldsymbol{P}}^{\mu}\epsilon
-\frac{i}{2}\sqrt{X_{{\rm M}}^2}\, \Gamma_{\circ}\Gamma_{\mu\nu}
\epsilon[\tilde{\boldsymbol{X}}^{\mu},\tilde{\boldsymbol{X}}^{\nu}]\bigr).
%, \\\{\bar{\epsilon}{\cal Q}, X_{\circ}^{\mu}\}_{{\rm D}}&=-\frac{1}{2\sqrt{-P_{\circ}^2}}{\rm Tr}(\bar{\epsilon}\boldsymbol{V}P_-\Gamma_{\circ}\tilde{\Gamma}^{\mu}\boldsymbol{\Theta})
\end{align}
The algebra of supercharge is 
\begin{align}
\{\bar{\epsilon}_1{\cal Q}, \bar{\epsilon}_2{\cal Q}\}_{{\rm D}}
&=-2(\bar{\epsilon_1}\Gamma_{\circ}\epsilon_2)
{\rm Tr}\Bigl(
\frac{1}{2}\tilde{\hat{\boldsymbol{P}}}^2-
\frac{1}{4}X_{{\rm M}}^2[\tilde{\boldsymbol{X}}^{\mu}, \tilde{\boldsymbol{X}}^{\nu}][\tilde{\boldsymbol{X}}_{\mu}, \tilde{\boldsymbol{X}}_{\nu}]\nonumber \\
&
+\frac{i}{2}\sqrt{X_{{\rm M}}^2}(\bar{\boldsymbol{\Theta}}\Gamma_{\circ}\Gamma_{\mu}[\boldsymbol{X}^{\mu}, \boldsymbol{\Theta}])\Bigr)\nonumber \\
&+2(\bar{\epsilon}_1\Gamma_{\circ}\Gamma_{\mu}\epsilon_2)\sqrt{X_{{\rm M}}^2}{\rm Tr}
\bigl(
i\tilde{\boldsymbol{X}}^{\mu}[ \tilde{\boldsymbol{X}}^{\nu},\tilde{\boldsymbol{P}}_{\nu}]-\frac{1}{2}i\tilde{\boldsymbol{X}}^{\mu}[\boldsymbol{\Theta}, \Gamma^0\Gamma_{\circ}\boldsymbol{\Theta}]_+\bigr).
\label{susyalgebra}
\end{align}
This result can be regarded just to be a covariantized version of the 
results well-known in the light-front Matrix theory. In our context, this  shows that,  since the first part on the r.h.side is proportional to 
the effective squared mass of this system 
up to a field-dependent SU($N$) gauge transformation 
exhibited in the second part, 
the commutator $[\delta_{\epsilon_1},\delta_{\epsilon_2}]$ 
induces an infinitesimal translation $s\rightarrow s
-2\bar{\epsilon}_1\Gamma_{\circ}\epsilon_2$ of the 
invariant proper-time parameter $s$. This of course reflects the fact 
that the dynamical supersymmetry is associated with the 
internal dynamics of this system. On the other hand, 
the supersymmetry, represented by $\delta_{\varepsilon}$, of the center-of-mass system 
does not induce the translation of 
the proper time: instead, it directly induces the translation 
of the center-of-mass coordinate $X_{\circ}^{\mu}$ without 
any shift of the proper time parameter. 
Because of this, it is appropriately called to be 
``kinematical" supersymmetry. A similar nature of 
the composition of kinematical and dynamical 
supersymmetries had already been apparent in 
the light-like formulation. It becomes more 
evident in our covariant formulation, due to manifestly 
different roles played by 
the internal proper time parameter and by the coordinates of target spacetime. 

That the matrix gauge fields are transformed 
in realizing dynamical supersymmetry is related to the fact 
that the mass-shell condition must be understood 
in conjunction with the Gauss law constraints, 
as we have emphasized in the purely bosonic case. 
The full action gives the following expressions for 
Gauss constraints of matrix type, apart from \eqref{bgauss}: 
\begin{align}
&\boldsymbol{G}_{A}\equiv i[\boldsymbol{X}^{\mu}, \boldsymbol{P}_{\mu}]-
\frac{i}{2}[\boldsymbol{\Theta}, \Gamma^0\Gamma_{\circ}\boldsymbol{\Theta}]_+\approx 0, 
\label{gauss1}\\
&\boldsymbol{G}_{B}\equiv X_{{\rm M}}\cdot \hat{\boldsymbol{P}}\approx 0, 
\label{gauss2}\\
&\boldsymbol{G}_{Z}\equiv P_{\circ}\cdot \hat{\boldsymbol{X}}=0, \quad 
P_{\circ}\cdot \hat{\boldsymbol{P}}=0
\end{align}
where in the last line we also wrote down the constraint 
derived as the equation of motion for the auxiliary field 
$\boldsymbol{K}$ in the gauge $\boldsymbol{K}=0$. 
It is easily seen that all of these constraints are 
invariant under supersymmetry transformation:
\[
\delta_{\epsilon}\boldsymbol{G}_A=0, \quad 
\delta_{\epsilon}\boldsymbol{G}_B=0, 
\quad %\delta_{\epsilon}(P_{\circ}\cdot \hat{\boldsymbol{X}})=0, 
\delta_{\epsilon}\boldsymbol{G}_Z=0, 
\quad 
\delta_{\epsilon}(P_{\circ}\cdot \hat{\boldsymbol{P}})=0. 
\]
On the other hand, 
the squared mass is 
\begin{align}
{\cal M}^2&\approx N{\rm Tr}(\hat{\boldsymbol{P}}\cdot \hat{\boldsymbol{P}}) 
-\frac{N}{6}\bigl<
[X^{\mu},X^{\nu}, X^{\sigma}], [
X_{\mu},X_{\nu},X_{\sigma}]\nonumber \\
&+i\frac{N}{2}\bigl<\bar{\Theta}, \Gamma_{\mu\nu}[
X^{\mu},X^{\nu}, \Theta]\bigr>\nonumber \\
&=N{\rm Tr}\Bigl(
\hat{\boldsymbol{P}}\cdot 
\hat{\boldsymbol{P}}- \frac{1}{2}\bigl(
X_{{\rm M}}^2
[\boldsymbol{X}^{\nu},\boldsymbol{X}^{\sigma}][\boldsymbol{X}_{\nu},\boldsymbol{X}_{\sigma}]\nonumber \\
&-2[X_{{\rm M}}\cdot \boldsymbol{X}, \boldsymbol{X}^{\nu}]
[X_{{\rm M}}\cdot\boldsymbol{X}, \boldsymbol{X}_{\nu}]\bigr)
+i\bar{\boldsymbol{\Theta}}\Gamma_{\mu\nu}
X_{{\rm M}}^{\mu}[\boldsymbol{X}^{\nu}, \boldsymbol{\Theta}]
\Bigr)
\end{align}
which is an equality under the 
above Gauss constraints, and is not itself 
invariant against the dynamical super transformations, 
satisfying 
\begin{align}
%\delta_{\epsilon}\Bigl(\frac{1}{e}{\cal H}\Bigr)= 
\delta_{\epsilon}\Bigl({\rm Tr}(\boldsymbol{A}\boldsymbol{G}_A-
\hat{\boldsymbol{B}}\boldsymbol{G}_B+\boldsymbol{Z}
\boldsymbol{G}_{Z}
%P_{\circ}\cdot\hat{\boldsymbol{X}}
)
-\frac{1}{2N}{\cal M}^2\Bigr)=0. 
\end{align}
It is therefore indispensable to take into account the 
Gauss constraints in treating the mass-shell condition, 
which itself is invariant against both 
kinematical and dynamical supersymmetry transformations. 
Of course, the invariance of the mass-shell condition 
under the kinematical supersymmetry is 
ensured by 
$\delta_{\varepsilon}P_{\circ}^{\mu}=\delta_{\varepsilon}X_{{\rm M}}^{\mu}=0$. 

In the light-front gauge, the mass-shell condition reduces to
\begin{align}
&P_{\circ}^2+{\cal M}_{{\rm lf}}^2\approx 0\nonumber \\
&{\cal M}_{{\rm lf}}^2\equiv 
N{\rm Tr}\bigl(
\hat{\boldsymbol{P}}^i\hat{\boldsymbol{P}}^i
-\frac{1}{2}X_{\rm M}^2 [
\boldsymbol{X}_i,\boldsymbol{X}_j][
\boldsymbol{X}_i,\boldsymbol{X}_j]
+i\sqrt{X_{{\rm M}}^2}\boldsymbol{\Theta}
\Gamma_i[\boldsymbol{X}_i, \boldsymbol{\Theta}]
\bigr).\nonumber 
\end{align}
Here we made a rescaling of the fermion matrix 
in order to fit it into the usual normalization of the 
light-front matrix theory,
\begin{align}
\boldsymbol{\Theta}\rightarrow \sqrt{2}\Bigl(-\frac{P^+_{\circ}}{P_{\circ}^-}\Bigr)^{-1/4}\boldsymbol{\Theta}. 
\end{align}
By repeating the procedure of deriving effective action in 
the bosonic case, we obtain the following effective action 
for the light-front theory in the first-order form:
\begin{align}
A_{{\rm lf}}=
\int ds\, \Bigl[{\rm Tr}\Bigl(
\hat{\boldsymbol{P}}_i\frac{D\hat{\boldsymbol{X}}_i}{Ds}
+\frac{1}{2}\boldsymbol{\Theta}
\frac{D\boldsymbol{\Theta}}{Ds}\Bigr)
-\frac{1}{2N}{\cal M}_{{\rm lf}}^2
\Bigr].
\end{align}
In the case of the IMF gauge, the corresponding result is 
\begin{align}
&A_{{\rm spat}}=\int ds \Bigl[{\rm Tr}\Bigl(\hat{\boldsymbol{P}}_i\frac{D\hat{\boldsymbol{X}}_i}{Ds}
+\frac{1}{2}\boldsymbol{\Theta}
\frac{D\boldsymbol{\Theta}}{Ds}\Bigr)
-P_{\circ}^0\Bigr], \\
&P_{\circ}^0=
\sqrt{
(P_{\circ}^{10})^2+{\cal M}_{{\rm lf}}^2
}.
\end{align}

\section{Conclusions}
In this lecture, I have proposed a re-formulation of Matrix theory in such a way that  
full 11-dimensional covariance is manifest, on the basis of the DLCQ 
interpretation of the light-front Matrix theory. 
It is successfully shown that the latter is obtained 
by a gauge-fixing of higher gauge symmetries from a covariant theory. 
The higher gauge symmetries are established in the framework of 
a Lorentz covariant canonical formalism, by 
starting from Nambu's generalization of the 
ordinary Hamilton mechanics. 

From the viewpoint of full 11-dimensional formulation of 
M-theory, the present work is not yet complete, as we will 
discuss shortly. However, I hope that this construction 
would be as an intermediate step toward our 
ultimate objective of constructing M-theory. 

The problems left unsolved include the followings, among many others. 
\begin{enumerate}
\item Dynamics of Matrix theory:

It remains, for instance, to see whether 11-dimensional matrices 
and the associated M-variables can provide any new insight for 
representing various currents and conserved (and topological) 
charges in the large $N$ limit. It is also 
worthwhile 
to study various scattering problems of graiton-partons 
in a manifestly covariant fashion by quantizing 
the present system using covariant gauges.

\item Background dependence and/or independence:

How to extend the present formulation to include non-trivial 
backgrounds, especially, curved background spacetimes? 
This is not straightforward, due to the intrinsic non-locality and 
novel higher gauge symmetries of the present model. 
Perhaps resolution of this problem would require full 
quantum mechanical treatments, remembering that 
interactions of subsystems, such as gravitons, are 
loop effects of off-diagonal matrix elements. 

\item Covariant re-formulation of the Matrix string theory:

There is a closedly
 related cousin to Matrix theory: 
the so-called Matrix string theory.\cite{dvv} The latter can be 
regarded as a natural matrix regularization\cite{sy}  
of supermembrane theory when the membranes are wrapped 
along the compactified circle. 
It may be possible to extend the present formulation to 
this case too. If successfully done, it may provide us a 
new method of dealing with second-quantized strings 
in a manifestly covariant fashion in (9,1) dimensions. 

\item Anti D0-branes:

This is one of the most pressing but difficult issues remaining. 
To include anti D0-branes, the SU($N$) gauge symmetry 
must be extended to the product of at least 
two independent gauge structure with SU($N$)$\times$ SU($M$). 
Furthermore, corresponding to the pair creation and 
annihilation of D0-anti D0 pairs, it should be possible to 
describe dynamically processes with varying $N$ and $M$ 
but keeping $N-M$ conserved. 
In other words, such a theory should be formulated in a Fock space\cite{yone2} 
with respect to the sizes of matrices. This is a very difficult issue,  
to which appropriate attention has not been paid yet. 

Related to this problem is that the dynamical supersymmetry 
is expected in general to be spontaneously broken when 
D0 and anti D0 coexist.\cite{sen2}\cite{hyo} Then, we should expect that 
even the dynamical supersymmetry would be 
realized in an intrinsically non-linear fashion. 
From this point of view, the present formulation of 
supersymmetry must be regarded as still tentative. 
It might be necessary to extend the bosonic 
higher gauge symmetry to include higher fermionic 
gauge symmetry, which may be a 
counterpart of the $\kappa$-symmetry of classical 
supermembranes, such that our projection 
condition for fermionic variables is regarded as a gauge-fixing 
condition for such higher fermionic 
gauge symmetry. 

\end{enumerate}
%I hope to come back to some of these issues in future.

%\newpage
\noindent
{\bf Acknowledgements} 

I would like to thank the organizers of the workshop 
for giving me 
the opportunity of presenting this lecture, and 
also for providing enjoyable atmosphere and hospitality 
during the workshop. 

The present work is supported in part by Grant-in-Aid for 
Scientific Research (No. 25287049) from the Ministry of 
Educationl, Science, and Culture.

\end{document}